\documentclass{emulateapj}

\usepackage{natbib}
\usepackage{graphicx}
\usepackage{txfonts}

\slugcomment{\it Accepted by ApJ - May 30, 2007}%

\shorttitle{Infrared SEDs of $z \sim$0.7 Star-Forming Galaxies}
\shortauthors{Zheng et al.}

\begin{document}


%________________________________________________________________
\title{Infrared Spectral Energy Distributions of $z\sim 0.7$ Star-Forming Galaxies}


%________________________________________________________________
\author{Xian Zhong\ Zheng\altaffilmark{1,2}, Herv\'e Dole\altaffilmark{3}, 
  Eric F.\ Bell\altaffilmark{1}, Emeric\ Le Floc'h\altaffilmark{4},
  George H.\ Rieke\altaffilmark{5}, Hans-Walter\ Rix\altaffilmark{1} and David\ Schiminovich\altaffilmark{6}} 
\altaffiltext{1} {Max-Planck Institut f\"ur Astronomie, K\"onigstuhl 17,
  D-69117 Heidelberg, Germany} 
\altaffiltext{2} {Purple Mountain Observatory, Beijing-West Road 2, Nanjing 210008, P. R. China; xzzheng@pmo.ac.cn} 
\altaffiltext{3} {Institut d'Astrophysique Spatiale (IAS), bat 121, F-91405 Orsay (France); Universit\'e Paris-Sud 11 and CNRS (UMR 8617)}
\altaffiltext{4} {Spitzer Fellow, Institute for Astronomy, University of Hawaii, 2680 Woodlawn Drive, Honolulu, HI 96822}
 \altaffiltext{5} {Steward Observatory, University of Arizona, 933 N Cherry
  Ave, Tucson, AZ 85721} 
\altaffiltext{6}{Department of Astronomy, Columbia University, New York, NY 100
27}

\begin{abstract}

  We analyze the infrared (IR) spectral energy distributions (SEDs) for
  $10\micron < \lambda_{\rm rest} < 100\micron$ for $\sim 600$ galaxies at
  $z \sim 0.7$ in the extended Chandra Deep Field South by stacking their
  {\it Spitzer} 24, 70 and 160\,$\micron$ images.  
  We place interesting constraints on the average IR SED {\it shape} in two
  bins: the brightest 25\% of $z \sim 0.7$ galaxies 
  detected at 24\,$\micron$, and the remaining 75\% of 
  individually-detected galaxies.  Galaxies without individual detections 
  at 24\,$\micron$ were not well-detected at 70\,$\micron$ and 160\,$\micron$
  even through stacking.
  We find that the average IR SEDs of $z\sim 0.7$ star-forming galaxies
  fall within the diversity of $z\sim 0$ templates.  
  While dust obscuration $L_{\rm IR}/L_{\rm UV}$ seems to be only a function
  of star formation rate (SFR; $\sim L_{\rm IR}+L_{\rm UV}$), not of redshift,
  the dust temperature of star-forming galaxies (with SFR $\sim 10$\,M$_{\sun}\,{\rm yr}^{-1}$) at a given IR luminosity was lower at $z \sim 0.7$ than today.
  We suggest an interpretation of this phenomenology 
  in terms of dust geometry: intensely star-forming galaxies at $z\sim 0$
  are typically interacting, and
  host dense centrally-concentrated bursts of star formation and
  warm dust temperatures.  At $z \sim 0.7$, the bulk of intensely
  star-forming galaxies are relatively undisturbed spirals and irregulars, 
  and we postulate
  that they have large amounts of widespread lower-density star formation, 
  yielding lower dust temperatures for a given IR luminosity.
  We recommend what IR SEDs are most suitable for modeling
  intermediate redshift galaxies with different SFRs. 

\end{abstract}


%%%%%%%%%%%%%%%%%%%%%%
\section{Introduction}

Dusty, intensely star-forming galaxies (SFR$ >10$\,$M_\odot$\,yr$^{-1}$) are the dominant contributors to the $z \geq 0.5$
cosmic SFR density \citep[e.g.,][]{Flores99,Elbaz02,Pozzi04,LeFloch05}. 
The thermal dust emission from these galaxies accounts for
much of the cosmic IR background, which contains half of the
radiation energy of the extragalactic background light
\citep{Hauser01,Lagache05,Dole06}. 
Therefore understanding the IR SEDs of the dusty star-forming galaxies is essential for
mapping the evolution of star formation as a function of 
cosmic time, and is a key observational ingredient of our understanding
of galaxy evolution.

Detailed studies of the IR SEDs over the full range of 3-1000\,$\micron$
have only been carried out for nearby galaxies \citep[e.g.,][]{Dale05}, finding that the IR SED shape is 
correlated with dust temperature \citep{Soifer91,Dale02,Chapman03,Lagache03}.  
Interestingly, despite this diversity 
of SED shapes, local galaxies' mid-IR (10-30\,$\micron$)
luminosities are tightly correlated with their total IR luminosities 
with a scatter of $\sim 0.3$\,dex \citep{Chary01,Papovich02,Takeuchi05b,Dale05}.

This {\it local} observation has often been used as 
a key assumption in studies exploiting 
deep mid-IR imaging from the {\it Infrared Space 
Observatory} at 15\,$\micron$ and the {\it Spitzer Space
Telescope} at 24\,$\micron$ \citep[e.g.,][]{Flores99,Flores04,Zheng04,Hammer05,Bell05,Melbourne05}.  Such studies
have found significantly enhanced star formation at $0.5 \la z \la 1$, 
compared to the present day.  Yet, this conclusion rests critically
on the extent to which mid-IR luminosities reflect the total IR luminosity.  
Unfortunately, only a small fraction of mid-IR detected
sources can be  individually detected in the far-IR bands
(e.g., {\it Spitzer} 70 \& 160\,$\micron$), owing to limited signal-to-noise
ratio (S/N) and source confusion \citep{Dole04b,Frayer06}.  
Thus, testing of this key assumption remains limited to the brightest
sources \citep[e.g.,][]{Sajina06,Borys06} or is indirect \citep[e.g.,][]{Appleton04,Yan05,Pope06,Marcillac06}. 

The goal of this paper is to explore the average IR SEDs of a 
stellar mass-limited sample
$z \sim 0.7$ galaxies in the extended Chandra Deep Field-South (E-CDFS). 
In previous works, we have shown that stacking noise-limited
images of a set of galaxies allows one to securely detect the mean flux of the
galaxy set substantially below the individual detection limit \citep{Zheng06,Dole06}. 
Here, we stack at longer wavelengths, i.e., 70\,$\micron$ and 
160\,$\micron$, to empirically determine the population-averaged
IR SED.  We combine these results with morphologies, and average 
fluxes at shorter wavelengths, to place constraints on the 
dust extinction and SEDs of these galaxies.
In \S 2 we describe the multi-wavelength data used to construct galaxy SEDs
and the samples of $z\sim 0.7$ galaxies. \S 3 
presents our stacking methods. In particular we test the results of stacking
noise and confusion limited 70 and 160\,$\micron$ images. 
In \S 4 we present the properties of the average SEDs.  
Discussion and conclusion are given in \S 5. Throughout this paper,
we assume $\Omega_M$\,=\,0.3, $\Omega_{\Lambda}$\,=\,0.7 and
$H_0$\,=\,70\,km\,s$^{-1}$\,Mpc$^{-1}$ for a $\Lambda$-CDM cosmology.
All magnitudes are given in the Vega system except where otherwise specified.

%%%%%%%%%%%%%%%%%%%%%%%%%%%%%
\section{The data and samples}

\subsection{The data}

We use {\it Spitzer} 24, 70 and 160\,$\micron$ data to study the
thermal dust emission of $z \sim 0.7$ galaxies. 
In addition, we include deep ultraviolet data from the
{\it Galaxy Evolution Explorer} (GALEX: \citealt{Martin05a}),
optical data from the Classifying Objects by Medium-Band Observations
(COMBO-17; \citealt{Wolf03}) survey,
and four band (3.6, 4.5, 5.8 and 8.0\,$\micron$) Infrared
Array Camera (IRAC; \citealt{Fazio04}) data to construct the
stellar SED of a galaxy.  
These data cover wavelength range from 0.15\,$\micron$ to
160\,$\micron$ in the observed frame, equal to the rest-frame range 
$\sim$0.09 to $\sim$100\,$\micron$ for $z=0.7$. 

The COMBO-17 survey has imaged the $30\farcm5\times 30\arcmin$ E-CDFS in
five broad ($U, B, V, R$ and $I$) and 12 medium optical bands, providing 
high-quality astrometry  (uncertainties $\sim 0\farcs 1$) based on a very deep
$R$-band image (26\,mag at the 5$\sigma$ limit), photometric
redshifts ($\delta z/(1+z) \sim 0.02$ at $m_{\rm R}<23$; \citealt{Wolf04}) and
stellar masses \citep{Borch06}
for $\sim 11,000$ galaxies with $m_{\rm R}<24$. We use the photometric
redshift and stellar mass catalogs to select galaxy samples.

GALEX ultraviolet 
observations provided deep far-ultraviolet (FUV; 1350--1750\AA) and
near-ultraviolet (NUV; 1750--2800\AA) images centered on the E-CDFS. 
The FUV and NUV images have a field of view of one square degree, a typical 
point-spread function (PSF) of full width at half-maximum (FWHM) $\sim5\arcsec$, a resolution of 1$\farcs$5\,pixel$^{-1}$ and
a depth of 3.63\,$\mu$Jy at the 5$\sigma$ level. 
The data reduction and source detection is described in
\citet{Morrissey05}.

The deep IRAC 3.6, 4.5, 5.8 and 8.0\,$\micron$ imaging data and MIPS 24, 70 and
160\,$\micron$ imaging data were obtained as part of the first run of MIPS GTO
observations \citep{Rieke04}. A rectangular field of $\sim
90\arcmin\times30\arcmin$ was observed in all bands (with small shifts between
different bands). The effective exposure time is 500\,s for the four IRAC
band images, 1378\,s\,pix$^{-1}$ for MIPS 24\,$\micron$, 
600\,s\,pix$^{-1}$ for 70\,$\micron$ and
120\,s\,pix$^{-1}$ for 160\,$\micron$.
IRAC 3.6 and 4.5\,$\micron$ images have a PSF of FWHM $\sim 1\farcs$8
and 5.8 and 8.0\,$\micron$ images have a PSF of FWHM $\sim 2\farcs$0 
\citep{Huang04}.
The 24\,$\micron$ image has a PSF of FWHM $\simeq$6$\arcsec$. 
Sources are detected at 24\,$\micron$ down to 83\,$\mu$Jy (at 80\% completeness; see \citealt{Papovich04} for details of data reduction, source detection
and photometry). The 70\,$\micron$ image is characterized by a PSF of
FWHM\,$\simeq$\,18$\arcsec$ and a resolution of 9$\farcs$9\,pixel$^{-1}$. 
The 160\,$\micron$ image has a PSF of FWHM\,$\simeq$\,40$\arcsec$ and a
resolution 
of 16$\arcsec$\,pixel$^{-1}$. Sources with fluxes of $f_{70}>$15\,mJy
can be individually resolved at 70\,$\micron$ and of $f_{160} >
50$\,mJy at 160\,$\micron$ \citep[see][for details]{Dole04a}.

We take COMBO-17 astrometry as the reference coordinate and
cross-correlate all other band catalogs with the COMBO-17 catalog. 
In each cross-correlation, we use bright stars and compact sources to estimate
the systematic offsets and uncertainties between two coordinates. 
A position tolerance of 4$\sigma$ uncertainty is adopted so that objects in
the two catalogs having coincident coordinates within the tolerance 
(corrected for the systematic offsets) are identified as the same objects. 
The nearest COMBO-17 object is chosen if multiple ones exist within the 
tolerance. 
The adopted tolerances are 2$\farcs$5 (FUV), 3$\farcs$0 (NUV), 
% 0$\farcs$9 ($J$ and $K$)
1$\farcs$0 (3.6\,$\micron$), 1$\farcs$2 (4.5\,$\micron$), 1$\farcs$5
(5.8 and 8.0\,$\micron$) and 2$\farcs$2 (24\,$\micron$). For the 70 and
160\,$\micron$ catalogs,  we firstly cross-correlate them 
with the 24\,$\micron$ catalog as individually-detected 70 and 160\,$\micron$
sources are bright at 24\,$\micron$; we then associate 70 and
160\,$\micron$ objects with COMBO-17 objects using the coordinates of the
corresponding 24\,$\micron$ sources.  The tolerance between 70/160
and 24\,$\micron$ is 5$\arcsec$/16$\arcsec$ respectively.

\begin{figure}[] \centering 
  \includegraphics[width=0.48\textwidth,clip]{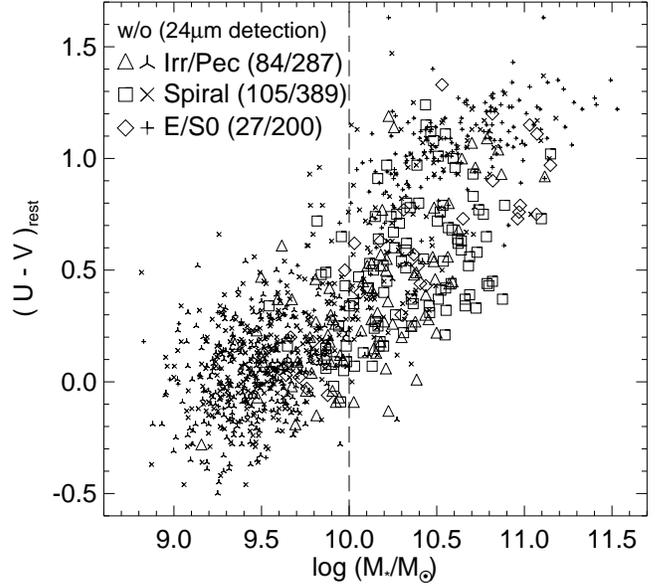}
\caption{The rest-frame color $U-V$ versus stellar mass for a 
sample of 1092 galaxies of known (GEMS) morphology in the redshift slice $z=0.7\pm0.05$. 
Open symbols show the objects detected individually at 24\,$\micron$
($f_{24}>83\mu$Jy) and skeletal symbols represent individually-undetected
ones. Most red ($U-V >$0.7) elliptical/lenticular 
galaxies and blue ($U-V <$0.7) small ($M_\ast <$10\,$M_\odot$)
spiral/irregular/peculiar galaxies are not individually detected at
24\,$\micron$. The number of objects in each category is given in the brackets.
The {\it dashed} line shows the mass cut $M_\ast = 10^{10}$\,$M_\odot$ that we
apply in the subsequent analysis.}\label{mass_UV}
\end{figure}

\begin{figure*}[] \centering 
  \includegraphics[width=0.3\textwidth,clip]{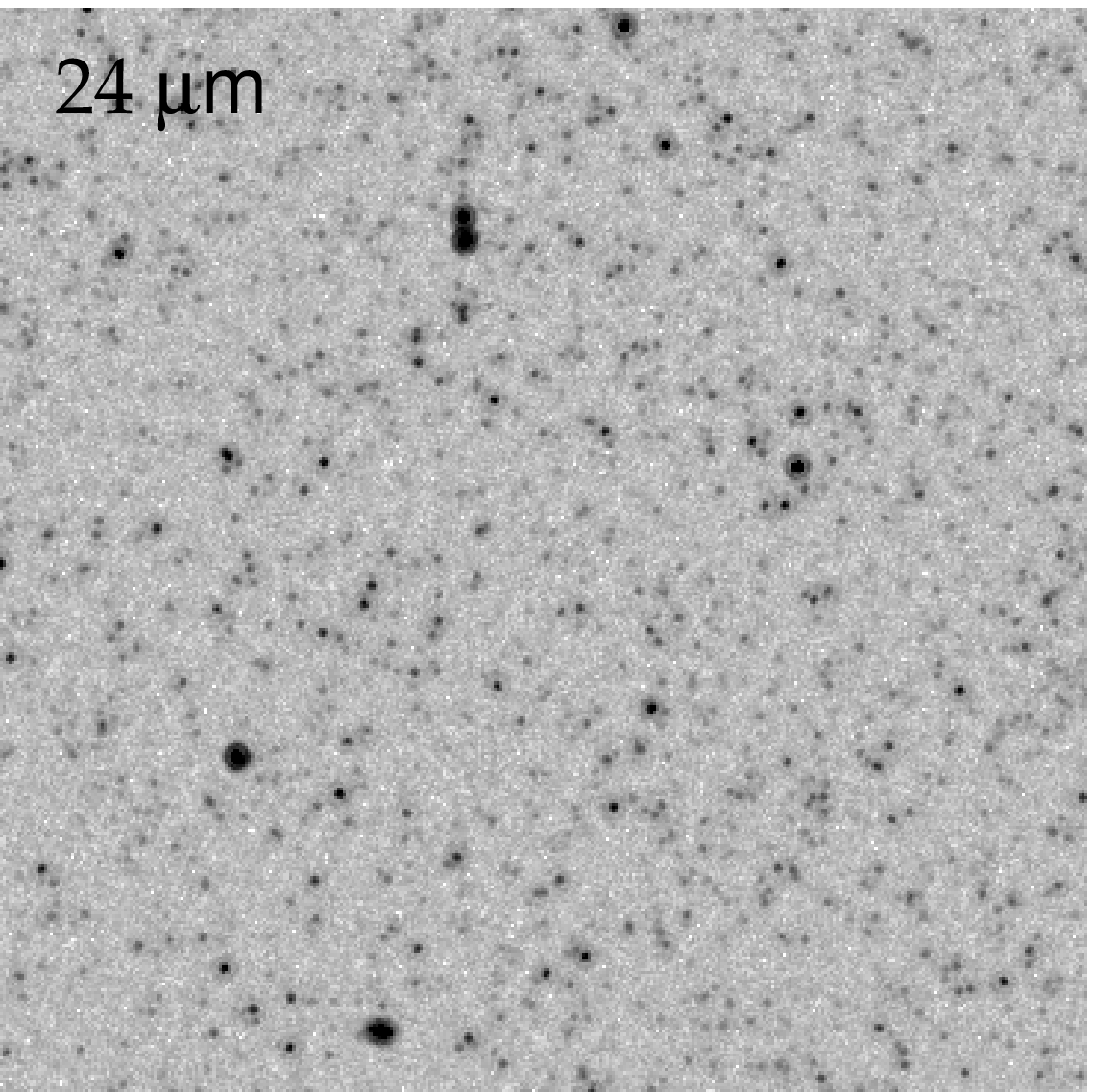}
  \includegraphics[width=0.3\textwidth,clip]{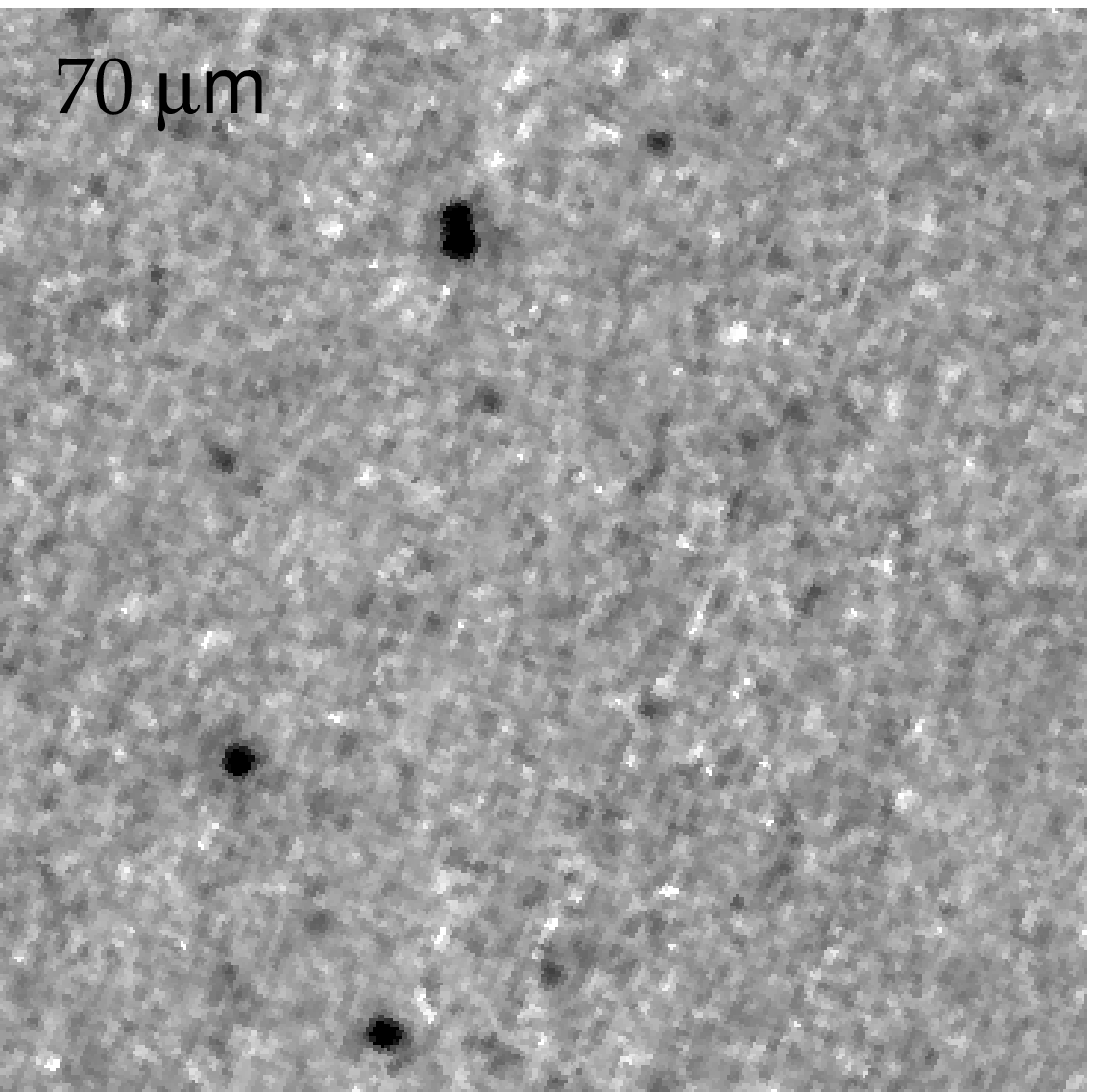}
  \includegraphics[width=0.3\textwidth,clip]{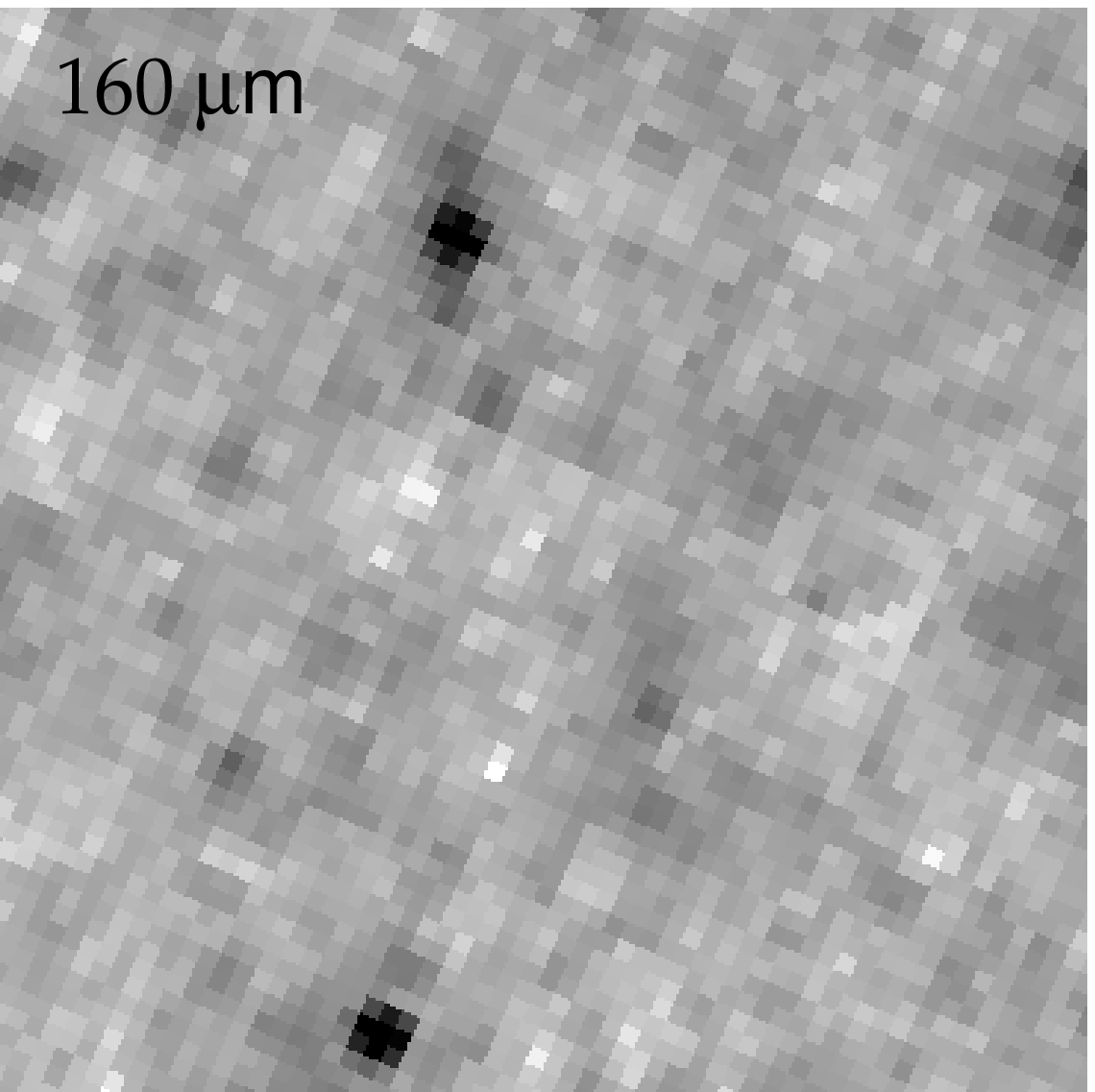}
  \parbox[t]{0.3\textwidth}{\hskip 1cm }
  \includegraphics[width=0.3\textwidth,clip]{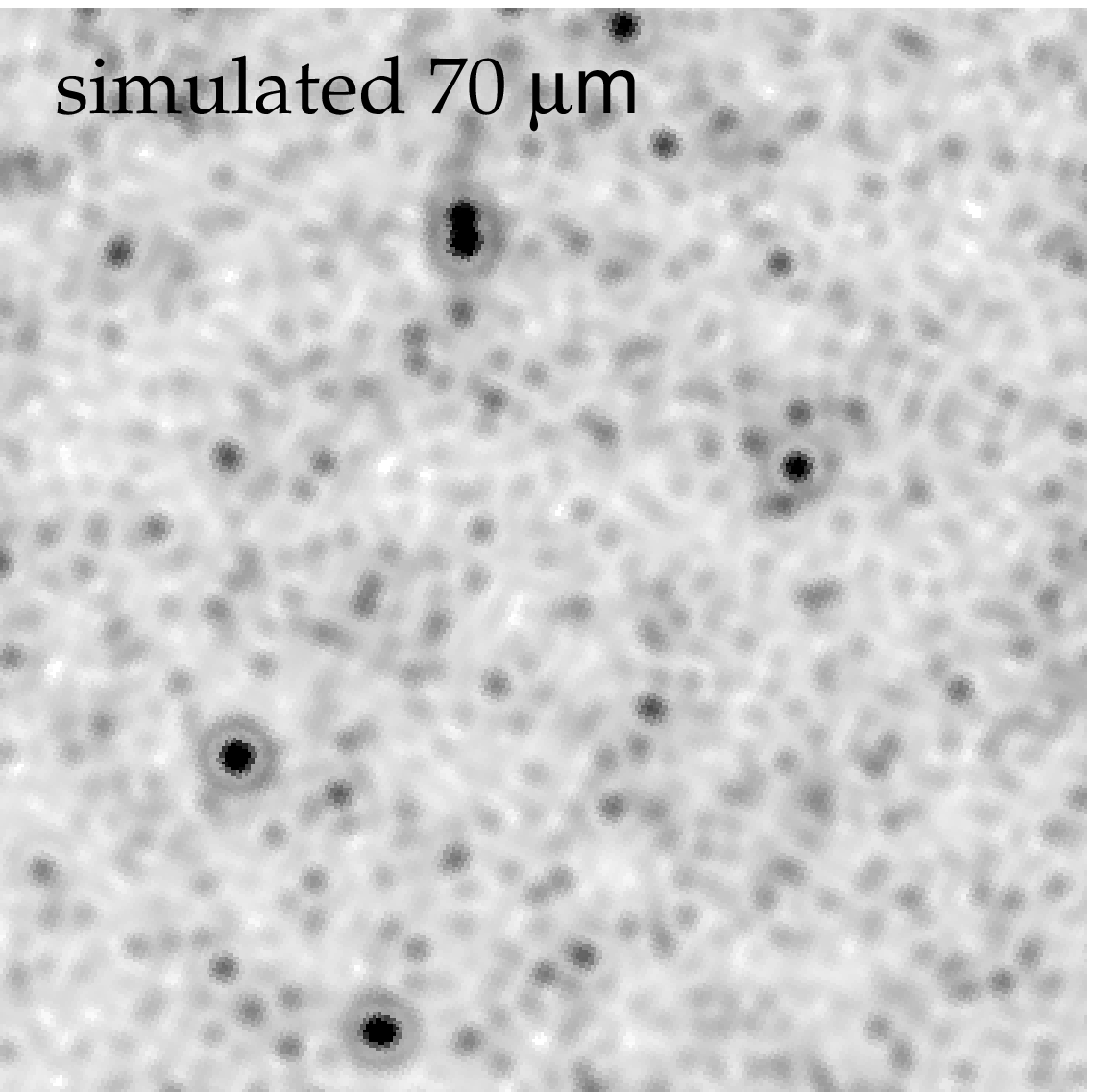}
  \includegraphics[width=0.3\textwidth,clip]{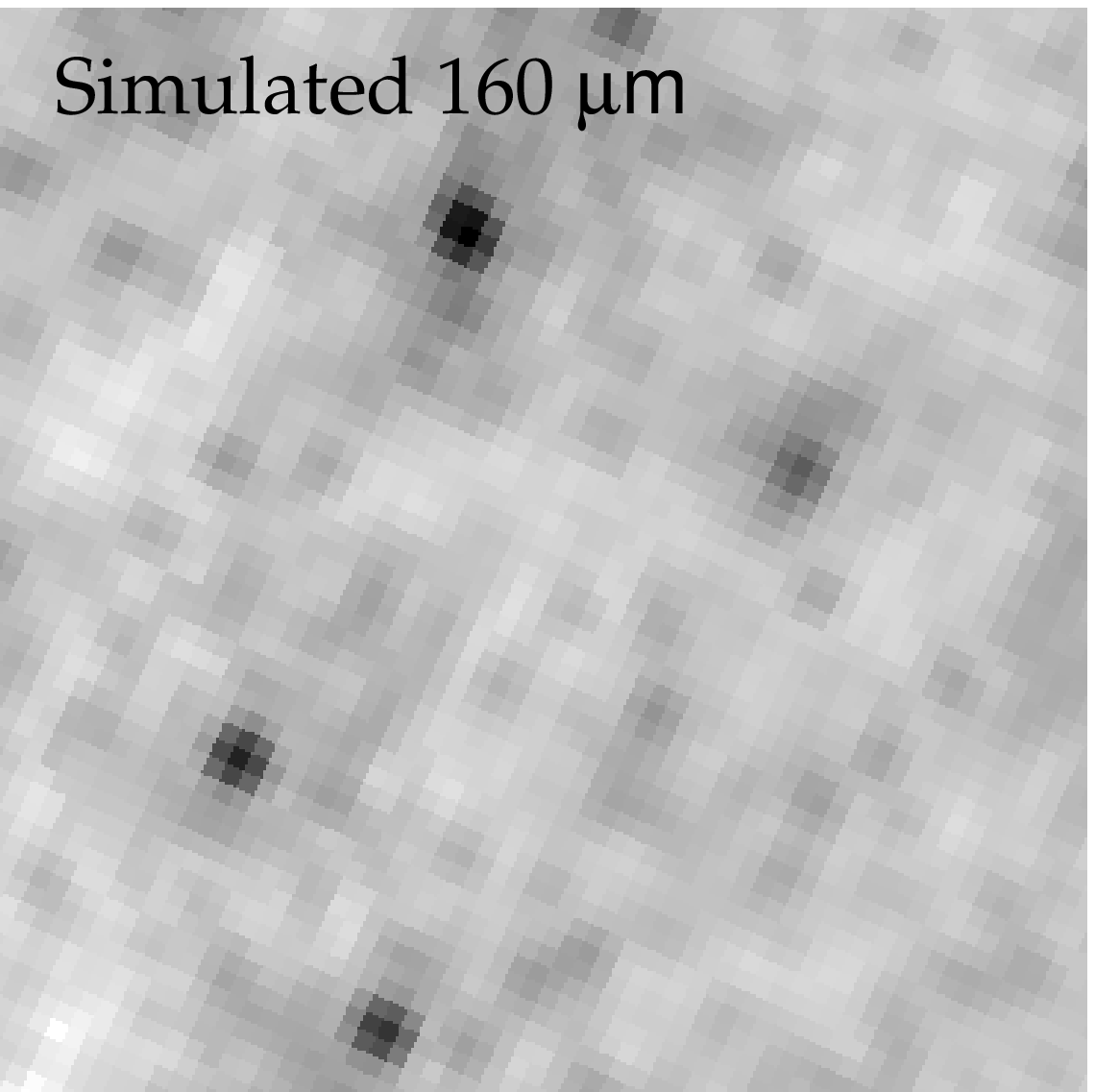}
\caption{Observed MIPS 24, 70 and 160\,$\micron$ image sections (15$\arcmin\times15\arcmin$) in the extended CDFS
  and simulated 70 and 160\,$\micron$  images.
  The simulated 70 and 160\,$\micron$ images are created by
  degrading the 24\,$\micron$ image to the 70 and 160\,$\micron$
  resolution.  No additional noise is added to the simulated
  images.}\label{images}
\end{figure*}

\subsection{The samples}\label{samples}

We combine spectroscopic redshifts from the VLT VIMOS Deep Survey \citep{LeFevre05} and the GOODS survey \citep{Vanzella05,Vanzella06} with photometric redshifts
from the COMBO-17 survey \citep{Wolf04} to select sample
galaxies. 
The E-CDFS has HST imaging from the Galaxy Evolution from Morphology and SEDs
\citep[GEMS;]{Rix04} Survey. Visually classified morphologies
are available for 1458 galaxies with $m_{\rm R}<24$ in a thin redshift slice
$z=0.7\pm0.05$ \citep{Bell05}. 
Of the 1458 galaxies, 1114 galaxies have all observations,
including GALEX,  IRAC and MIPS observations. 
To avoid contamination from active galactic nuclei (AGNs), we remove 22 X-ray detected sources
in the Chandra 250\,ks observation \citep{Lehmer05}, 
leaving a sample of 1092 galaxies. Spectroscopic redshifts are available for 64\% of the sample galaxies. 
The contribution from the X-ray-undetected AGNs to the total 24\,$\micron$ luminosity of $z<1$ galaxies is suggested to be $<\sim 10$\% \citep{Zheng06,Brand06}. This will not have significant effects on our results.

The sample demographics are shown in Figure\ \ref{mass_UV}.  The 
sample is limited by $R$-band apparent magnitude ($m_{\rm R}<24$), 
corresponding to approximately the rest-frame $B$-band at $z \sim 0.7$.
Accordingly, the completeness of the sample, in terms 
of stellar mass, is a strong function of color: the 
mass limit for red (old or dusty) galaxies is 
$M_\ast \sim 10^{10}$\,$M_\odot$ whereas blue
galaxies can be included down to almost
$M_\ast \sim 10^{9}$\,$M_\odot$.  We choose to impose
a stellar mass cut of $M_\ast \geq  10^{10}$\,$M_\odot$ in 
what follows; not only are almost all 24\,$\micron$ emitters 
above this mass cut, but also the completeness of this
sample is not a strong function of color (i.e., age or dust 
obscuration). We also select another sample to explore the relationship between the IR SED shape and 24\,$\micron$ luminosity; we extend the redshift slice to  0.6$<z<$0.8 to increase the number of sample galaxies. The final sample then comprises
some 579 galaxies with $M_\ast \ge 10^{10}$\,$M_\odot$ in the redshift
range 0.6$<z<$0.8. Of them, 218 are individually-detected at 
24\,$\micron$ with fluxes in excess of $>83\mu$Jy; none is individually
detected at 70 or 160\,$\micron$. 
X-ray detected sources have been excluded.

\section{Image stacking}

As outlined earlier, current missions are unable to yield
individual detections for the vast majority of intermediate-redshift
objects at far-IR wavelengths, owing both to 
contributions from instrumental noise and confusion noise.
In order to place constraints on the shape of the IR 
SEDs of `typical' star-forming intermediate redshift
galaxies, stacking on the positions of known star-forming
galaxies can lower the effective noise \citep{Zheng06}, allowing 
detection of the `average' galaxy.  In \citet{Zheng06}
we presented a description of stacking of 24\,$\micron$
data (in that case to resolve the 24\,$\micron$ luminosity
of dwarf galaxies); we  briefly summarize the most important aspects of 24\,$\micron$
stacking in \S \ref{stack:24}.  The focus of this paper is stacking
at longer wavelengths, at 70\,$\micron$ (PSF FWHM 18$\arcsec$) and 
160\,$\micron$ (FWHM 40$\arcsec$), discussed in \S \ref{stack:70160}.

\subsection{FUV, NUV and MIPS 24\,$\micron$ image stacking}\label{stack:24}

While much of this paper describes stacking results for subsamples
that are individually detected at 24\,$\micron$, some subsamples
are not individually detected at 24\,$\micron$.  Furthermore, 
the 24\,$\micron$, FUV and NUV images share many of the same characteristics:
the PSFs have similar FWHM, and at each wavelength, $\ga 1/2$
of the extragalactic background at that wavelength is 
resolved by these images (i.e., the images are only 
mildly confusion-limited; e.g., \citealt{Papovich04,Xu05,Dole06}).  
Accordingly, stacking of 
FUV, NUV and 24\,$\micron$ images is carried out in the same
way \citep[see][for more details]{Zheng06}.

Three basic steps are adopted to derive the mean fluxes for galaxy subsets.  
First, we subtract all individually-detected sources from the 
images. This is done using the software tool STARFINDER\footnote{STARFINDER
  gives identical results within the errors to the tool ALLSTAR in IRAF
  package.} \citep{Diolaiti00}
with an empirical PSF constructed from 18/42/56 bright point sources
at 24\,$\micron$/FUV/NUV respectively. 
Then we perform mean stack of the residual image postage stamps centered on
the optical coordinates of the objects that are individually undetected 
in the subset of interest. An aperture of 5$\arcsec$ is used to
integrate the 
central flux of the mean-stacked image and estimate the background from the
outer regions. Aperture corrections of a factor of 1.88/1.19/1.14, derived
from the empirical PSF at 24\,$\micron$/FUV/NUV respectively, are adopted 
to calibrate the stacked fluxes to the total fluxes. 
Last we sum the fluxes of individually-detected sources and the
stack flux of individually-undetected sources in each galaxy subset, 
giving the mean 24\,$\micron$/FUV/NUV fluxes. 
Uncertainties are derived from bootstrapping.

\begin{figure*}[] \centering 
  \includegraphics[width=0.49\textwidth,clip]{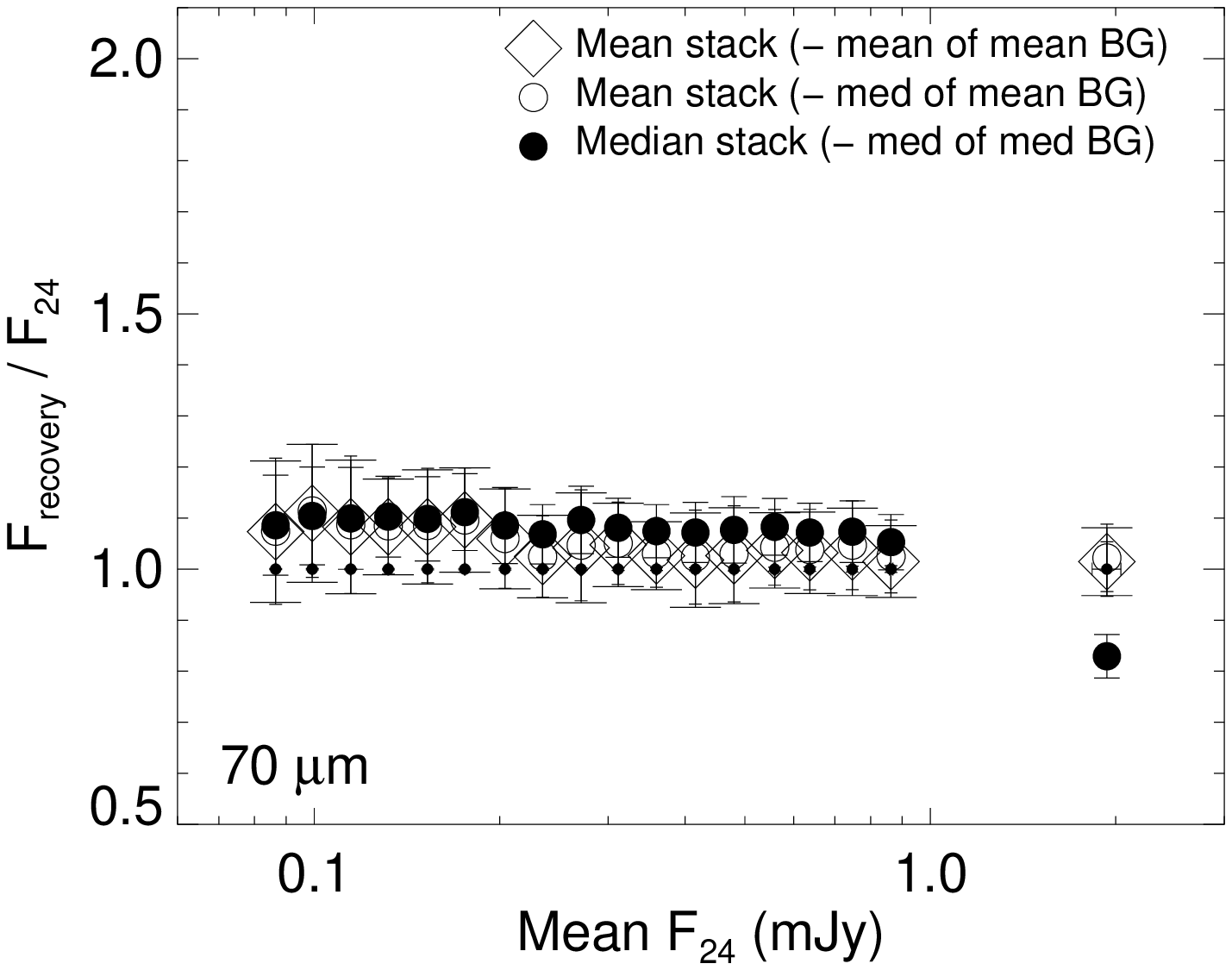}
  \includegraphics[width=0.49\textwidth,clip]{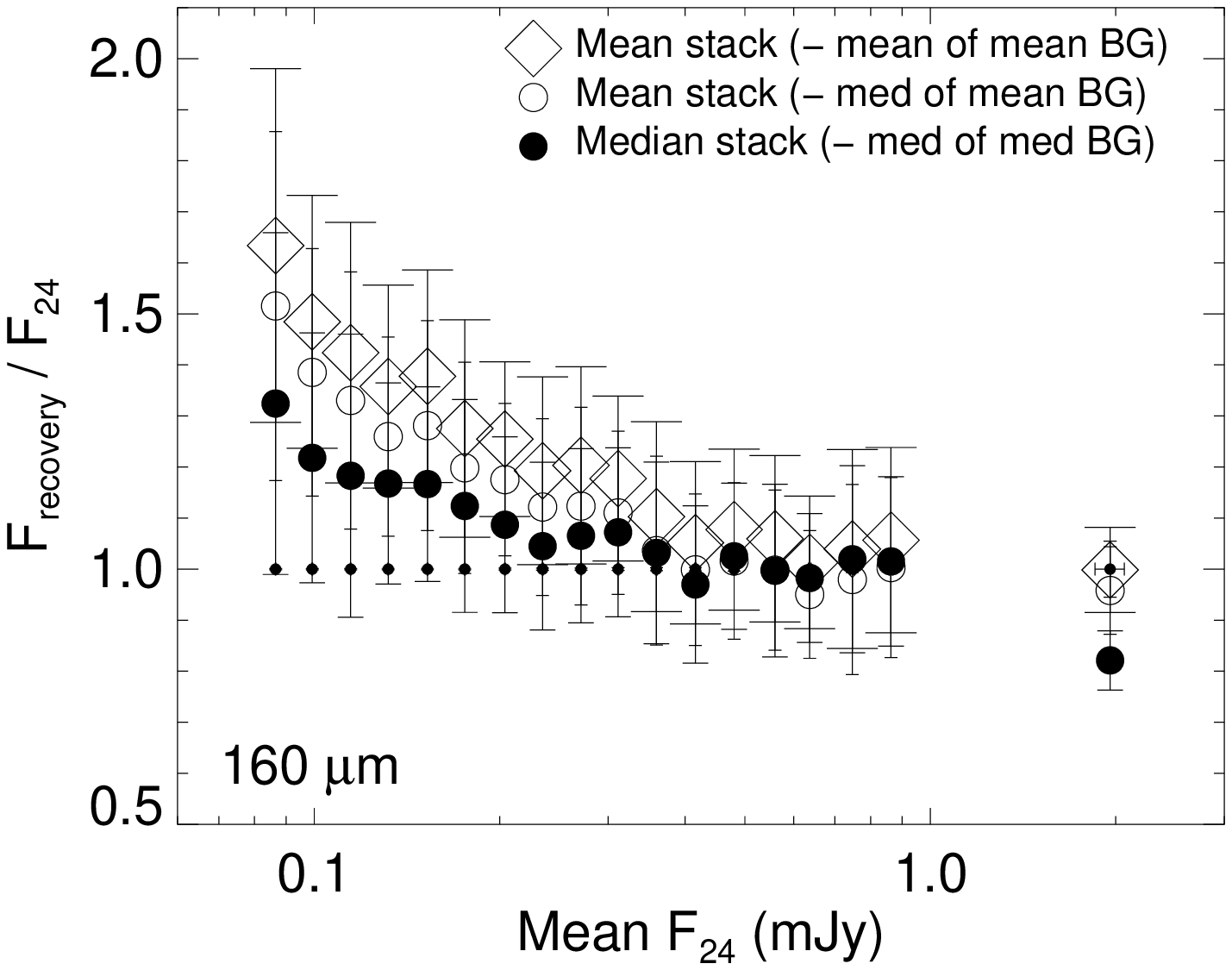}
\caption{Recovery of the mean fluxes of 
  {\it randomly} distributed 24\,$\micron$
  sources through stacking after convolution to the
  70\,$\micron$ ({\it left}) and 160\,$\micron$ ({\it right}) resolution.  
  The fluxes were taken from the 8255 individually-resolved sources in the 
  $\sim1\fdg 5\times0\fdg 5$ 24\,$\micron$ image and were split into 18 
  flux bins in the range from 83\,$\mu$Jy to 10\,mJy. The number of
  objects in each bin decreases with 
  increasing flux. Three measures of the stack
  flux are presented for each bin: the integrated 
central flux of the {\it mean} stack using the {\it mean} background for
sky subtraction; the integrated central flux of the {\it mean} stack using the
{\it median} background; and the integrated central flux of the {\it median}
stack using the {\it median} of the background (see text for details).
  Errorbars are derived from bootstrapping. While the flux can be recovered
  with little bias at 70\,$\micron$ resolution, flux recovery at
  160\,$\micron$  leads to a systematic overestimate. 
}\label{random}
\end{figure*}

\begin{deluxetable*}{ccccccccccccccccccc}
  \centering \tabletypesize{\scriptsize} \tablewidth{0pt}
  \tablecaption{The numbers of sources and average fluxes for
  18 flux bins composed of 8255 individually-resolved 24\,$\micron$ sources.
  \label{stackbin}}
\tablehead{No. & 1 & 2 & 3 & 4 & 5 & 6 & 7 & 8 & 9 & 10 & 11 & 12 & 13 & 14 & 15 & 16 & 17 & 18}
\startdata
$<f_{24}>$($\mu$Jy) & 87 & 99 & 115 & 132 & 153 & 176 & 204 & 235 & 271 & 311 & 359 & 416 & 480 & 559 & 637 & 747 & 863 & 1946 \\
Number & 1070 &  1259 &  1067 &  894 &  803 &  683 & 508 &  418 &  328 & 267 &  222 &  174 &  134 &  98 &  72 &  46 &  39 &  169
\enddata 
\end{deluxetable*}

\subsection{MIPS 70 and 160\,$\micron$ image stacking} \label{stack:70160}

The 70\,$\micron$ and 160\,$\micron$ PSFs are considerably larger than 
those at shorter wavelengths, yielding confused
images, resolving only the brightest, relatively nearby sources --- some 
$<$30\% of the extragalactic background at this wavelength \citep{Dole04a}.  
Galaxies at $0.6<z<0.8$ are heavily confused in all but the
brightest cases \citep{Dole04a,Lagache03,Lagache04}, requiring stacking
to gain insight into their long-wavelength IR SEDs \citep[see, e.g.,][]{Dole06}.  

The large angular extent of the long-wavelength PSFs
poses a significant challenge for those wishing to estimate
their average properties.  At $z$\,=\,0.7, the PSF size of 
the 70\,$\micron$ image 
(FWHM\,$\simeq 18\arcsec$) 
corresponds to a physical scale of $\sim$130\,kpc. For the 160\,$\micron$ image
(PSF FWHM\,$\simeq 40\arcsec$),  the corresponding scale is $\sim$290\,kpc.
Thus, the stacking results are a reflection of the IR-luminosity 
weighted two-point correlation function on $\ga 100$\,kpc scales, 
overestimating the true average fluxes of the galaxies of interest.  

In order to understand this source of systematic error 
in better detail, we carried out some simulations where
synthetic 24\,$\micron$ data (using the observed positions
and fluxes of individually-detected 24\,$\micron$ sources) 
are degraded in resolution to the resolution
of the 70\,$\micron$ and 160\,$\micron$ data.  For the purposes
of this test, we assume a constant ratio between 24\,$\micron$ flux
and the wavelength of interest.  We explore two cases.   First, 
the positions are randomly scrambled (i.e., the relative 
brightnesses of galaxies are preserved but their positions
are random).  This test gives an indication of the 
systematic effects of stacking randomly-positioned sources.
The second case is when both the fluxes and positions of
sources are preserved.  This second case is our `best'
estimate of the likely systematic uncertainties of stacking 
in a realistically-clustered case.  
Figure~\ref{images} shows the 24\,$\micron$ images degraded to the 70 and
160\,$\micron$ resolution respectively, compared with the 
observed MIPS 24, 70 and 160\,$\micron$ images. 
Because of confusion, only a handful of bright 24\,$\micron$ sources can
be individually resolved in the degraded images. 
Through stacking the mean flux can be estimated for 24\,$\micron$ source
subsets; the accuracy of the recovery of the mean flux 
shows how well stacking works at that corresponding image
resolution.

\begin{figure*}[] \centering
  \includegraphics[width=0.49\textwidth,clip]{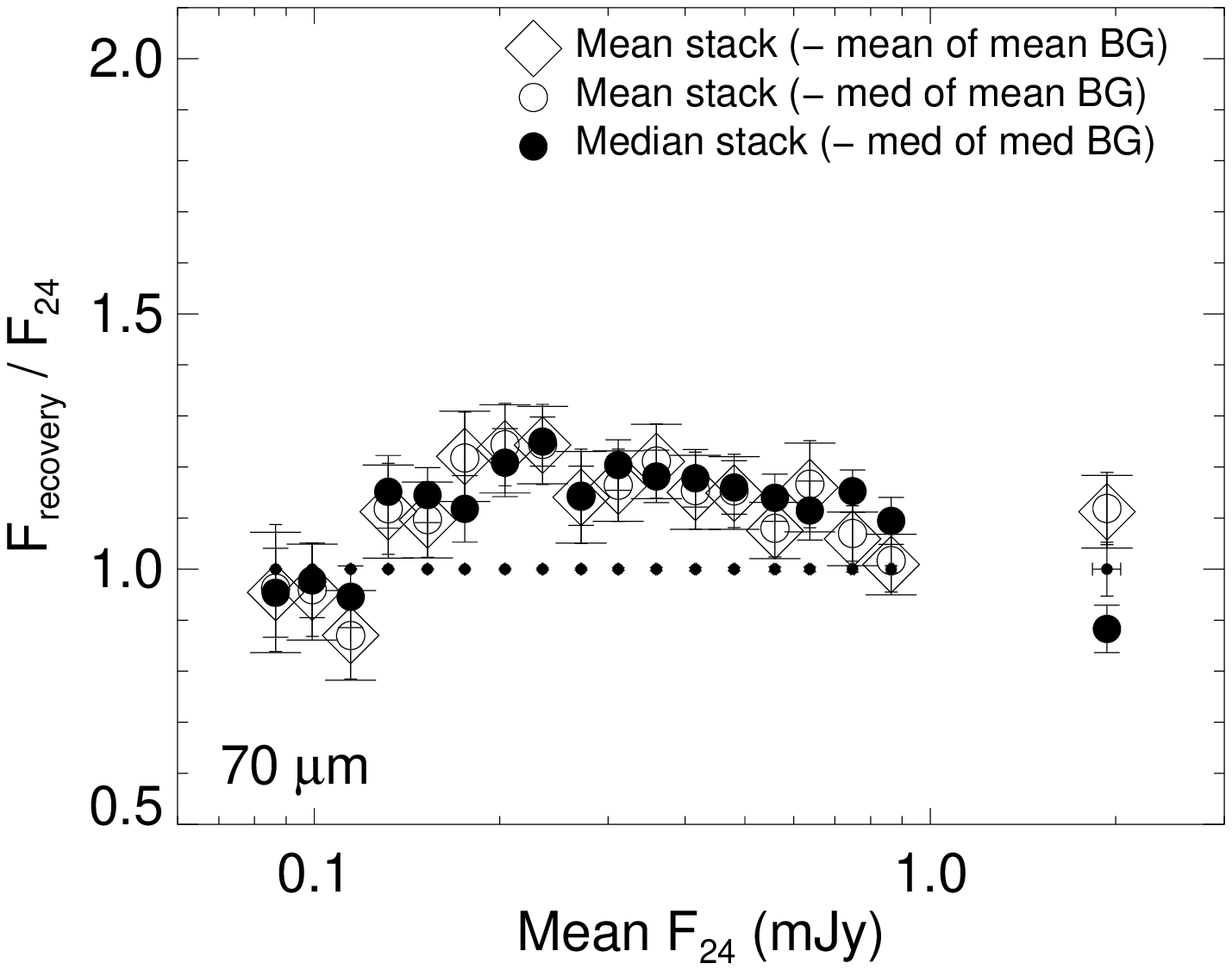}
  \includegraphics[width=0.49\textwidth,clip]{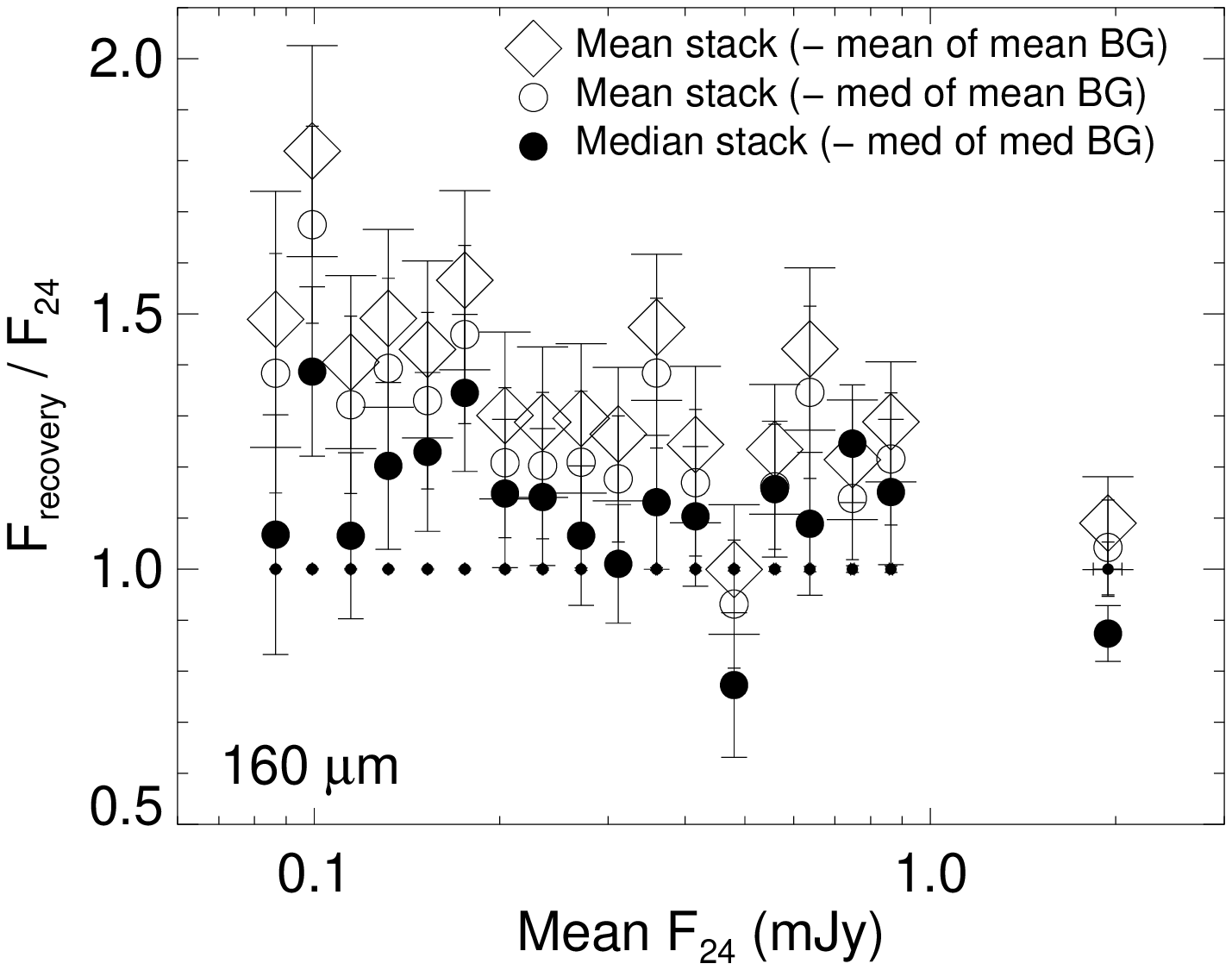}
\caption{The recovery of the mean properties of 
  24\,$\micron$-detected sources at their
  observed positions.  {\it Left}: Stacking results at
  70\,$\micron$ resolution. {\it Right}: Stacking results at 160\,$\micron$
  resolution.   The errorbars reflect the variance from bootstrapping.  
}\label{cdfs24}
\end{figure*}

\subsubsection{Stacking randomly distributed sources}\label{stack:random}

The whole 24\,$\micron$ mosaic image of the E-CDFS covers a
rectangular sky area of $\sim1\fdg5\times 0\fdg5$ and contains 8255
individually-detected sources at 5$\sigma$ detection limit (83\,$\mu$Jy).  
The vast majority of the 8255 sources ($f_{24}>83\mu$Jy) are point sources. 
Replacing all sources with the 24\,$\micron$ PSF (empirically constructed from
bright stars), we generated an artificial 24\,$\micron$ image having the 8255
sources {\it randomly} distributed into a $\sim1\fdg5\times 0\fdg5$ blank
field.  The artificial 24\,$\micron$ image was then 
degraded to the 70 and 160\,$\micron$ image resolution respectively.  
The degraded images are then stacked at the 24\,$\micron$ source positions, giving indications as to 
the expected quality of stacking results at 70\,$\micron$ and 160\,$\micron$.
We note that the replacement of the background 
blank field with the PSF-subtracted
24\,$\micron$ image (having some contribution from individually-undetected
sources) does not modify our results significantly in the
mean, adding only some modest additional scatter.

The 24\,$\micron$ sources were sorted into 18 bins in flux, ranging from
83\,$\mu$Jy to 10\,mJy \citep[see also][]{Dole06}, 
with $\sim 1000$ objects per bin at faint flux levels, 
and $\la 100$ per bin at brighter fluxes. The numbers of sources 
and average 24\,$\micron$ fluxes per bin are listed in Table~\ref{stackbin}.
In real stacks, 
we choose to PSF-subtract out the individually-detected sources, 
stacking the remaining image (in order to reduce bias from bright sources 
in background estimates).  In the real 70 and 160\,$\micron$
images, we detect $\sim 130$ sources in each image.  Thus, we must
subtract approximately this many bright sources from the simulated
images before stacking to avoid unrealistically biasing the results.
Ideally, one would subtract all detected sources in the simulated
70 and 160\,$\micron$ images; unfortunately, as the simulated
and real 70 and 160\,$\micron$ images have (unavoidably) somewhat
different noise and brightness distribution properties, the 
number of detected sources is different, and substantially higher
in the case of (the rather deeper) simulated 70\,$\micron$ data.  
Thus, we choose to PSF-subtract the 169 brightest sources (the brightest of
the 18 original bins) before stacking for the remaining 17 bins.

For each flux bin, postage stamps were cut 
from the simulated image centered on the
positions of the sources and were stacked. 
The size of postage stamps is 2$\farcm$5$\times$2$\farcm$5 for the
70\,$\micron$ image and 4$\arcmin \times 4\arcmin$ for the 160\,$\micron$
image, allowing for a proper background estimate. 
Two stacked images were created for each bin through averaging or medianing.
For the stacked 70\,$\micron$ postage stamp images, an aperture of radius
0$\farcm$49 is used to integrate the central stack flux and an annulus with
inner radius 0$\farcm$82 and outer radius 1$\farcm$23 to estimate the
background. For the stacked 160\,$\micron$ postage stamp images, the
corresponding aperture is of radius 1$\farcm$07 and the corresponding
annulus is of inner radius 1$\farcm$07 and outer radius 1$\farcm$87.
The mean and median of pixels in the annulus 
region were taken as the background
of the mean-stacked image for sky subtraction.  Only the median (nearly
identical to the mean) was adopted as the background of the median-stacked
image. Therefore, we derive three measures of the stack flux for each
subsample: the integrated 
central flux of the {\it mean} stack using the {\it mean} background for
sky subtraction; the integrated central flux of the {\it mean} stack using the
{\it median} background; and the integrated central flux of the {\it median}
stack using the {\it median} of the background.  
Finally, aperture corrections derived from model PSFs 
(http://ssc.spitzer.caltech.edu/mips/psffits/) were applied to correct the
estimates of stack flux to the total. The aperture correction is a factor of
1.30 for 70\,$\micron$,  1.25 for 160\,$\micron$ using the
median background, and 1.34 for 160\,$\micron$ using the mean
background. 

Figure~\ref{random} shows the results of stacking {\it randomly} distributed
sources at 70 and 160\,$\micron$ resolution. Uncertainties are derived from
bootstrapping. 
At 70\,$\micron$ resolution, the mean flux of 24\,$\micron$ sources of
comparable flux can be properly recovered over a flux range of one order
of magnitude.  At 160\,$\micron$ resolution, recovered fluxes 
are of reasonable quality at bright limits, and become progressively
more biased (by $\sim 20$-50\%) towards fainter limits.  
After some investigation, it became clear that the 
the stack recovery at 160\,$\micron$ is 
correlated with the number of objects in the stack bin.
The increase of objects in number leads to an increase of the overlap
between these objects within a fixed area. This increasing overlap 
leads to an increase of the stack flux relative to the input flux.
By stacking a number of identical sources without background and
foreground sources, we obtained similar results as shown in Figure~\ref{random}.
Indeed a point source contains 80\% of its flux in an area of 3.6 arcmin$^2$
in the 160\,$\micron$ imaging and an area of 0.9 arcmin$^2$ in the
70\,$\micron$ imaging.  
To fill in a $1\fdg5\times 0\fdg5$ field, it requires around 750
160\,$\micron$ sources or 3000 70\,$\micron$ sources, compared to $\sim$\,50 -
1260 objects in the stack bins.
Therefore the stack results at 70\,$\micron$  are little affected by the
overlap between the stack objects but the stack results at 
160\,$\micron$ are significantly influenced for stack bins of $\sim 1000$
objects. 

The three measures of the stack flux are nearly identical within the errorbars
($\sim$\,10\%) for simulated 70\,$\micron$ stacking 
and slightly different for simulated 160\,$\micron$ stacking, in
particular for the low flux bins, although the scatter is significant
($\sim$\,10 - 25\%).  Note that we divided sources into stack bins in terms of
their fluxes. For the stack bins having sources whose fluxes span a wide
range, the median stack 
substantially underestimates the mean flux.

\subsubsection{Stacking sources at their observed positions}

In order to build a more realistic picture of the expected uncertainties 
from stacking at 70\,$\micron$ and 160\,$\micron$, we repeat the last 
analysis except we keep both the fluxes and positions of the sources
fixed to those observed at 24\,$\micron$.  This gives insight into the 
influence of clustering on the results.  In this test, the observed
24\,$\micron$ map was degraded to the 70 and 160\,$\micron$
resolution respectively, as shown in Figure~\ref{images}. 
As previously, sources were split into 18 bins in 24\,$\micron$
flux, and the stacking results for each flux bin were calculated.
Figure~\ref{cdfs24} shows the results.  Comparing the results
with Figure~\ref{random}, one can clearly see that 
the mean and median recovered fluxes are significantly affected
by the clustering of sources on the sky.  
In general, our test suggests that stacking subsets of 70\,$\micron$
and 160\,$\micron$ sources may overestimate the mean flux somewhat; typical
biases are $\sim 20$\%, with uncertainties of order $\sim 20$\%.
Yet, the magnitude
and direction of these effects on different subsamples 
are difficult to estimate {\it a priori}.
Figure~\ref{cdfs24} demonstrates that
from bin to bin, the source clustering may lead to either an increase or a
decrease of the mean flux of the source subsample, depending on the actual
distribution and density of sources of different brightnesses on the sky.

\subsubsection{Stacking $z\sim$\,0.7 galaxies}\label{stackgal}

The above tests demonstrate that it will be difficult to 
estimate the mean flux at 70\,$\micron$ and 160\,$\micron$
to better than $\sim 20$\%, and that the extent of the bias towards higher
or lower flux is difficult
to predict with accuracy.  In this paper, we adopt a 
pragmatic approach: for each subsample that is stacked
at 70\,$\micron$ and 160\,$\micron$, we attempt to derive an individual
`bias correction' based on the 24\,$\micron$ image.  
Sources bright at 24\,$\micron$ are generally bright at 70 and
160\,$\micron$, although the 24\,$\micron$ to 70 or 160\,$\micron$
flux ratio varies from object to object at a scatter at the 0.5\,dex level \citep{Dale05}.   Thus, we adopt the simplistic
assumption that the 24\,$\micron$ image can be taken as a good proxy
for the 70 or 160\,$\micron$ image at 
$\simeq$\,6$\arcsec$ resolution and high S/N; 70\,$\micron$ and 160\,$\micron$
resolution images derived under this assumption (Fig.~\ref{images}) appear
a reasonable description of the real 70\,$\micron$ and 160\,$\micron$ images.
Thus, for a given subset of $z\sim 0.7$ galaxies, we estimated precisely the mean
24\,$\micron$ flux from their 24\,$\micron$ images. Then we degraded the
24\,$\micron$ images to 70 and 160\,$\micron$ image resolution. Two stack
fluxes were estimated by stacking each of the two sets of degraded images.  By
comparing the stack fluxes to the actual mean 24\,$\micron$ flux of the galaxy
subset, we obtained empirical corrections, which were applied to
the corresponding 70 and 160\,$\micron$ stack fluxes of the subset of
$z\sim 0.7$ galaxies, respectively.  Uncertainties in this correction 
are applied also, in quadrature, to the derived stacking results at 70\,$\micron$
and 160\,$\micron$.

\begin{figure}[] \centering 
  \includegraphics[width=0.48\textwidth,clip]{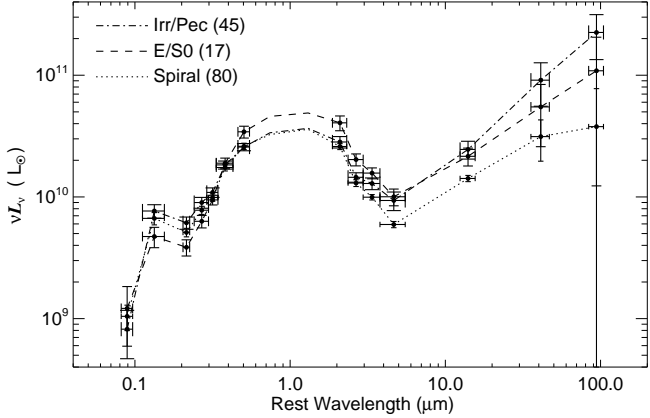}
\caption{The average SED, spanning a factor of 1000 in wavelength, as a function of morphological type for
  24\,$\micron$ detected ($f_{\rm 24}>83\mu$Jy) galaxies with $M_\ast \ge
  10^{10}$\,$M_\odot$ in a thin redshift slice 0.65$\leq z \leq$0.75. Vertical
  errorbars are the 1$\sigma$ uncertainties derived from bootstrapping and
  horizonal errorbars show the band widths. The number of objects in each
  morphology bin is given in the brackets.}\label{sedsir}
\end{figure}

\begin{figure}[] \centering 
  \includegraphics[width=0.48\textwidth,clip]{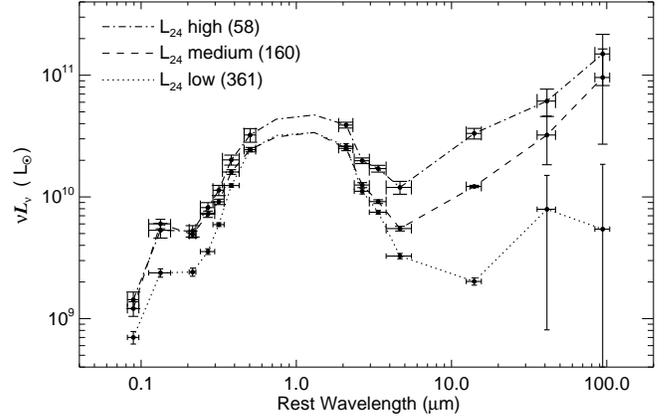}
\caption{Average SED as a function the 24\,$\micron$ luminosity for
  galaxies with $M^\ast\geq 10^{10}$\,$M_\odot$ in the redshift range
  0.6$<z<$0.8. The vertical errorbars show the 1$\sigma$ uncertainties
  and horizonal errorbars show the band widths. The 579 galaxies are split
  into three bins in terms of their 24\,$\micron$ 
  luminosities: $L_{24}$-high, $L_{24}$-medium and $L_{24}$-low. 
  The numbers in brackets denote the number of objects included in each bin.
}\label{sedsmass}
\end{figure}

\begin{deluxetable*}{lcccccccc}
  \centering \tabletypesize{\scriptsize} \tablewidth{0pt}
  \tablecaption{Average luminosities of massive-galaxy subsets in the redshift
  slice $0.6<z<0.8$. 
  \label{table}}

\tablehead{ N$_{\rm obj}$ & $\log <M^\ast>$ & $\log \nu L_{\nu,24}$ & $\log
  \nu  L_{\nu,70}^\mathrm{a}$  & corr.(70) & $\log \nu L_{\nu,160}^\mathrm{a}$  & corr.(160) & $\log <L_{\rm
  12--100}>^\mathrm{b}$ & $\log <L_{\rm IR}>^\mathrm{c}$ \\
  & ($M_\odot$) & ($L_\odot$) & ($L_\odot$) & & ($L_\odot$) & & ($L_\odot)$ & ($L_\odot$)}
\startdata
58  & 10.7 & 10.52$^{+0.04}_{-0.05}$ & 10.79$^{+0.10}_{-0.13}$ & 1.51$\pm$0.24 & 11.17$^{+0.16}_{-0.26}$ & 1.04$\pm$0.39 & 11.30$^{+0.10}_{-0.13}$  & 11.42$^{+0.07}_{-0.08}$ \\
160 & 10.5 & 10.09$^{+0.01}_{-0.01}$ & 10.51$^{+0.16}_{-0.24}$ & 1.03$\pm$0.32 & 10.98$^{+0.23}_{-0.55}$ & 0.92$\pm$0.61 & 11.04$^{+0.16}_{-0.27}$  & 11.12$^{+0.05}_{-0.05}$ \\
361 & 10.7 &  9.31$^{+0.03}_{-0.03}$ &  9.90$^{+0.28}_{-0.99}$ & 1.19$\pm$0.96 & $ <$10.27               & ...           & 10.18$^{+0.26}_{-0.75}$  & 10.28$^{+0.14}_{-0.21}$ \\
\enddata 
\tablenotetext{a}{The 70\,$\micron$ and 160\,$\micron$ luminosities
  estimated from stacking have included the corrections derived from degraded 24\,$\micron$ images. The corrections are listed in the table.}
\tablenotetext{b}{$L_{\rm 12--100}$ is the luminosity between 12 to
  100\,$\micron$, calculated by linearly interpolating observed 24, 70 and
  160\,$\micron$ luminosities (in logarithm space).}
\tablenotetext{c}{$L_{\rm IR}$ refers to the total IR luminosity between 8 to
  1000\,$\micron$. Local IR SED templates from \citet{Lagache04},\citet{Chary01} and \citet{Dale02} are used to fit the observed data points. The
  total IR luminosity is derived from the dust temperature-match
  templates. See text for details.}
\end{deluxetable*}

\section{Results}\label{results}

To explore the average IR SEDs of $z \sim 0.7$ star-forming galaxies, we first look into the dependence of the IR SED shape on galaxy morphology. Then, we investigate the relationship between the IR SED shape and the 24\,$\micron$ luminosity and the $z \sim 0.7$ IR SEDs to those of present-day star-forming galaxies. Finally, based on the IR SEDs, the extrapolation of the total IR luminosity from the 24\,$\micron$ luminosity is discussed.

\subsection{The relationship between IR SED and morphology}\label{consed}

We used a sample of 1092 galaxies of known morphology to investigate the
dependence of the IR SED shape on morphology. 
The 1092 galaxies are classified into three
morphological types: elliptical/lenticular (E/S0), spiral, and
irregular/peculiar. Figure~\ref{mass_UV} shows the relationship between the
rest-frame color $U-V$ and stellar mass. Objects detected at 24\,$\micron$ 
($f_{24}>83\mu$Jy) are marked with open symbols and
undetected ones with skeletal symbols. The detection limit of 83\,$\mu$Jy
corresponds to the monochromatic observed-frame 24\,$\micron$ luminosity  
$\nu L_{\nu,24} = 6\times 10^{10}$\,$L_\odot$ at $z=0.7$.  
The 24\,$\micron$-undetected galaxies are intrinsically faint in the IR bands
(see \S\ref{ir:24}), and we excluded them in constructing these 
composite SEDs.  To avoid selection bias in color (see \S\ref{samples}), we also excluded about one third of the 24\,$\micron$
detected galaxies with $M_\ast < 10^{10}$\,$M_\odot$, leaving  152
galaxies in the final sample.
We calculated the average luminosities in 14 bands for each 
subsample of 24\,$\micron$-detected galaxies: FUV \& NUV
from GALEX, $U,B,V,R$ and $I$ from the COMBO-17 survey, 
IRAC 3.6, 4.5, 5.8 and 8.0\,$\micron$, MIPS 24, 70 and
160\,$\micron$ bands.  The average luminosity of each 
band includes contributions from
both individually-detected and individually-undetected sources (at
wavelengths other than 24\,$\micron$).  
Errors in the average luminosities were derived from bootstrapping, including
contributions to the uncertainty from measurement errors in both the
individually-detected fluxes and the individually-undetected fluxes.
The average SED spans a range in the rest-frame of 0.1 to 100\,$\micron$. 

Figure~\ref{sedsir} shows the average SEDs of the three
subsamples.  Each of the three average SEDs is dominated by a
dust-extincted stellar spectrum at $\lambda<$\,5\,$\mu$m 
and emission by dust at $\lambda>$\,5\,$\mu$m.  
Irregular/peculiar galaxies show a higher ratio of dust to stellar
emission compared to the spirals and E/S0
galaxies. The slope of the IR SED (rest-frame
10-100\,$\micron$) is steeper for irregular/peculiar galaxies,
somewhat intermediate for E/S0 galaxies, and lowest for spirals although the
uncertainties of 70 and 160\,$\micron$ luminosities are large. 
The irregular/peculiar galaxies usually form stars in relatively
concentrated regions,  leading to a high star formation intensity (i.e. SFR per
unit area). In contrast, the spirals are often characterized by a
relatively low star formation intensity as the star-forming regions are
widely distributed over disks. The star formation
density for the E/S0 galaxies is somewhat between those of the
irregular/peculiar galaxies and spirals. The shapes of the IR SEDs are thus a
function of star formation intensity in the sense that the
dust temperature is primarily colder in systems of relatively lower star
formation intensity (we will return this topic in \S \ref{discuss}).

\subsection{The relationship between IR SED and 24\,$\micron$ luminosity} \label{ir:24}

We used a mass-limited sample to study the relationship between IR SED shape
and 24\,$\micron$ luminosity. This sample consists of 579 galaxies with
$M^\ast\geq 10^{10}$\,$M_\odot$ and  0.6\,$<z<$\,0.8. 
The sample galaxies were divided into three 24\,$\micron$
luminosity bins: $L_{24}$-high, $L_{24}$-medium and $L_{24}$-low. 
The first two bins contain 218 individually-detected 24\,$\micron$ sources
($f_{24}>$83\,$\mu$Jy) and all individually-undetected 24\,$\micron$ sources (361 of
the 579) are in the third bin. The $L_{24}$-high bin is chosen to
contain 58 brightest 24\,$\micron$ sources 
so that its total 24\,$\micron$ luminosity equals that of the
$L_{24}$-medium bin. Consequently the stacked 70 and 160\,$\micron$ fluxes are expected to have comparable signal-to-noise ratios for the two bins. Average luminosities in all 14 bands were
calculated for the three subsets of galaxies. 
The average luminosities in the 24, 70 and 160\,$\micron$ bands are listed in
Table~\ref{table}, along with the number of objects and mean stellar mass for
each of the three subsets. The empirical corrections adopted for the 70 and
160\,$\micron$ stack fluxes (see \S\ref{stackgal}) are also presented. 
Errors include the uncertainties in measurements and bootstrapping errors. 
Figure~\ref{sedsmass} shows the average SED from the rest-frame wavelengths 0.1
to 100\,$\micron$ as a a function of the 24\,$\micron$ luminosity. 
It is clear that the observed 24\,$\micron$ (rest-frame 14\,$\micron$)
luminosity is correlated with the 70 and 160\,$\micron$ IR luminosities for
massive galaxies at $z\sim 0.7$ in the sense that the 70 and 160\,$\micron$
IR luminosities increase as the 24\,$\micron$ luminosity increases; i.e., 24\,$\micron$
luminosity typically reflects high IR luminosity, rather than an enhanced
rest-frame mid-IR excess (see also Bavouzet et al. 2007, in preperation).
The $L_{24}$-high bin has an average galaxy stellar mass 0.2\,dex larger than
the $L_{24}$-medium bin; this can also be seen from the redder
optical colors and higher rest-frame $\sim 3$\,$\micron$ luminosity
of the $L_{24}$-high bin.
The $L_{24}$-low bin contains 361 massive galaxies that are individually
undetected at 24\,$\micron$, including early-type galaxies with little star
formation and late-type galaxies in the quiescent star formation phase (see 
Figure~\ref{mass_UV}).  The short-wavelength part of their average SED is dominated by a relatively 
old stellar population. The large errorbars of the 70 and 160\,$\micron$
luminosities compared to the small errorbar of the 24\,$\micron$ luminosity
are partially due to the intrinsic scatter among the sample galaxies in this
bin.

\begin{figure}[t] \centering 
  \includegraphics[width=0.48\textwidth,clip]{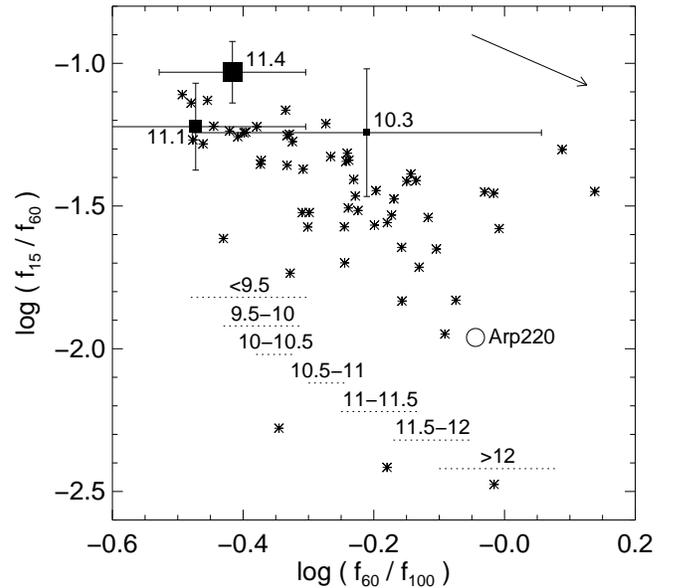}
\caption{Relative dust temperature estimate from the IR flux ratio $f_{60}/f_{100}$ versus $f_{15}/f_{60}$. 
  {\it Asterisks} represent local star-forming galaxies
  from \citet{Dale00} with IRAS observations at the 60 and
  100\,$\micron$ and ISO observation at 15\,$\micron$.  The ratio
  $f_{60}/f_{100}$  is an indicator of the dust temperature, which tends
  to be correlated with IR luminosity in the local Universe.  
  The {\it dotted lines} show typical values of $f_{60}/f_{100}$
  corresponding the given $\log (L_{\rm IR}/L_\odot)$ derived from
  IRAS data \citep{Soifer91}. These
  lines  are arbitrarily positioned on the Y-axis. 
The {\it open  circle} shows the well-known local ultraluminous IR galaxy
  Arp\,220.   The {\it squares} show the three subsets of massive  
  ($M^\ast\ge 10^{10}$\,$M_\odot$) galaxies in redshift slice 0.6$<z<$0.8,
  labeled with the corresponding IR luminosities 
  $\log (L_{\rm IR}/L_\odot)$. Note that errorbars of the two flux ratios
  are not independent. The arrow indicates the effect of an increase in
  $f_{60}$ by a factor of 1.5. 
}\label{colorcolor}
\end{figure}

\begin{figure*}[] \centering 
  \includegraphics[width=\textwidth,clip]{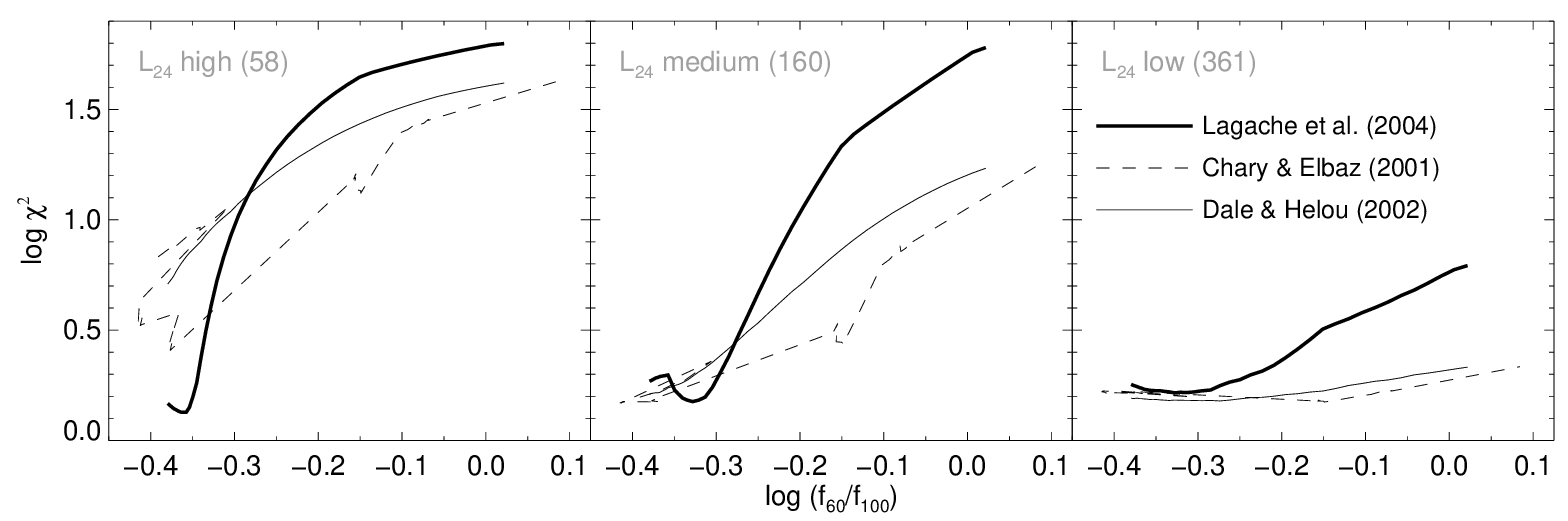}
  \includegraphics[width=\textwidth,clip]{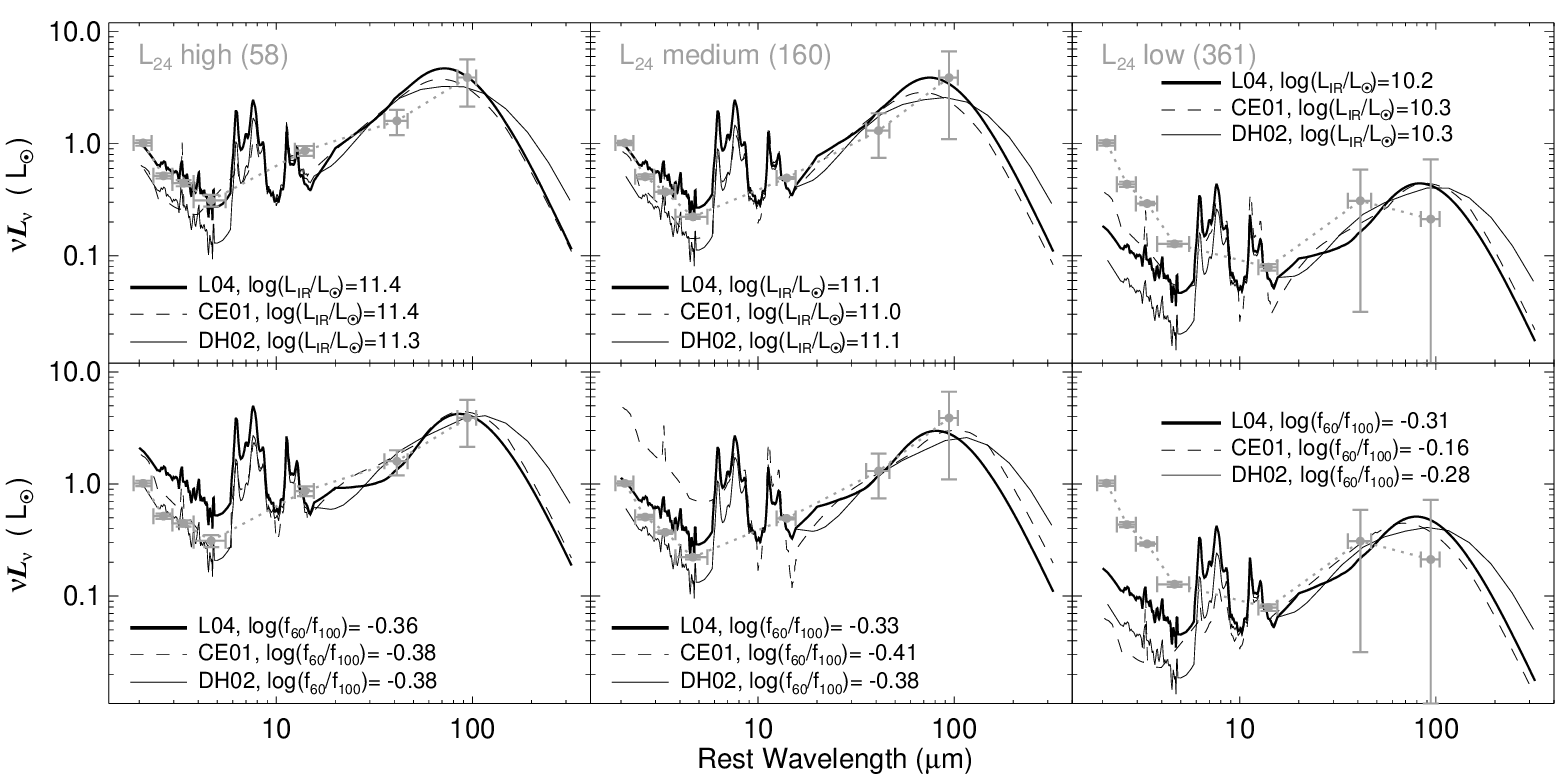}
\caption{Fitting three sets of local templates from \citet{Lagache04}, \citet{Chary01} and \citet{Dale02} to the average IR SEDs of three subsets of massive
  ($M^\ast\ge 10^{10}$\,$M_\odot$) galaxies in the redshift slice 0.6$<z<$0.8.  
  Only the three MIPS data points (24, 70 and 160\,$\micron$) of each observed IR
  SED are used in the fitting.  
  {\it Top}: The panels show $\chi^2$ as a function of the
  template's characteristic $\log (f_{60}/f_{100}$). 
  For each observed IR SED two best-fit templates are chosen: one is matched
  in luminosity (i.e., with the normalization constant equal to 
  unity) and the other is dust temperature match (or SED
  shape match; i.e., with the minimum $\chi^2$). 
  {\it Bottom}: The {\it upper panels} shows the best-fit luminosity match templates 
and the {\it lower panels} show the best-fit dust temperature match templates,
  compared to the observed SEDs from the rest-frame
  2 to 100\,$\micron$ delineated by IRAC and MIPS data points 
  for the $L_{24}$-high bin ({\it left panels}),
  the $L_{24}$-medium bin ({\it middle panels}) and the $L_{24}$-low bin ({\it
  right panels}). The SEDs are normalized to unity at 2.1\,$\micron$. 
  Note that the stellar components of the locoal templates are somewhat arbitrarily set. The disagreements between the averaged SEDs and the best-fit templates at $\lambda_{\rm rest} < 5\micron$ should be ignored.
}\label{sedsmodel}
\end{figure*}

\begin{figure*}[] \centering 
  \includegraphics[width=0.49\textwidth,clip]{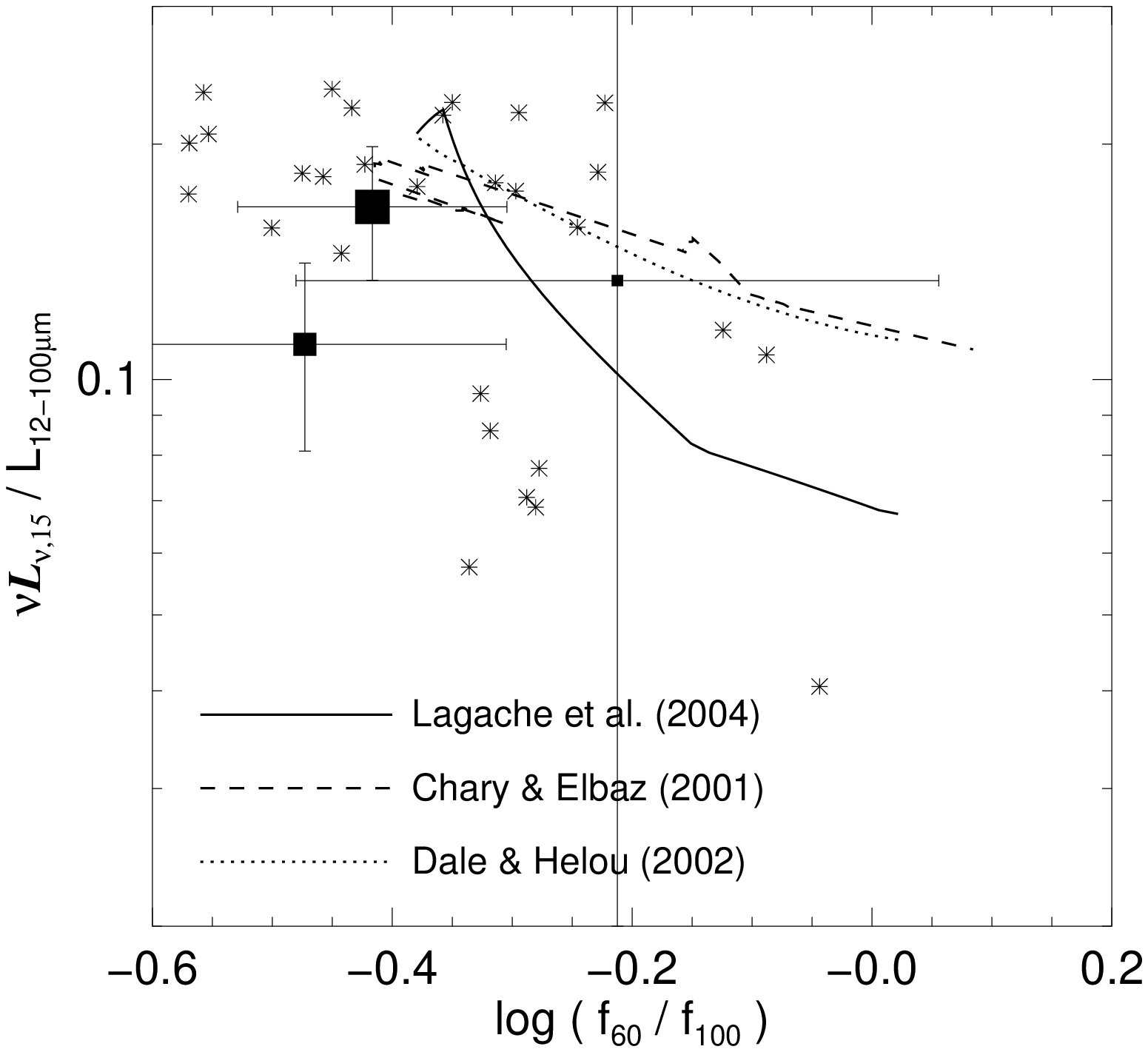}
  \includegraphics[width=0.49\textwidth,clip]{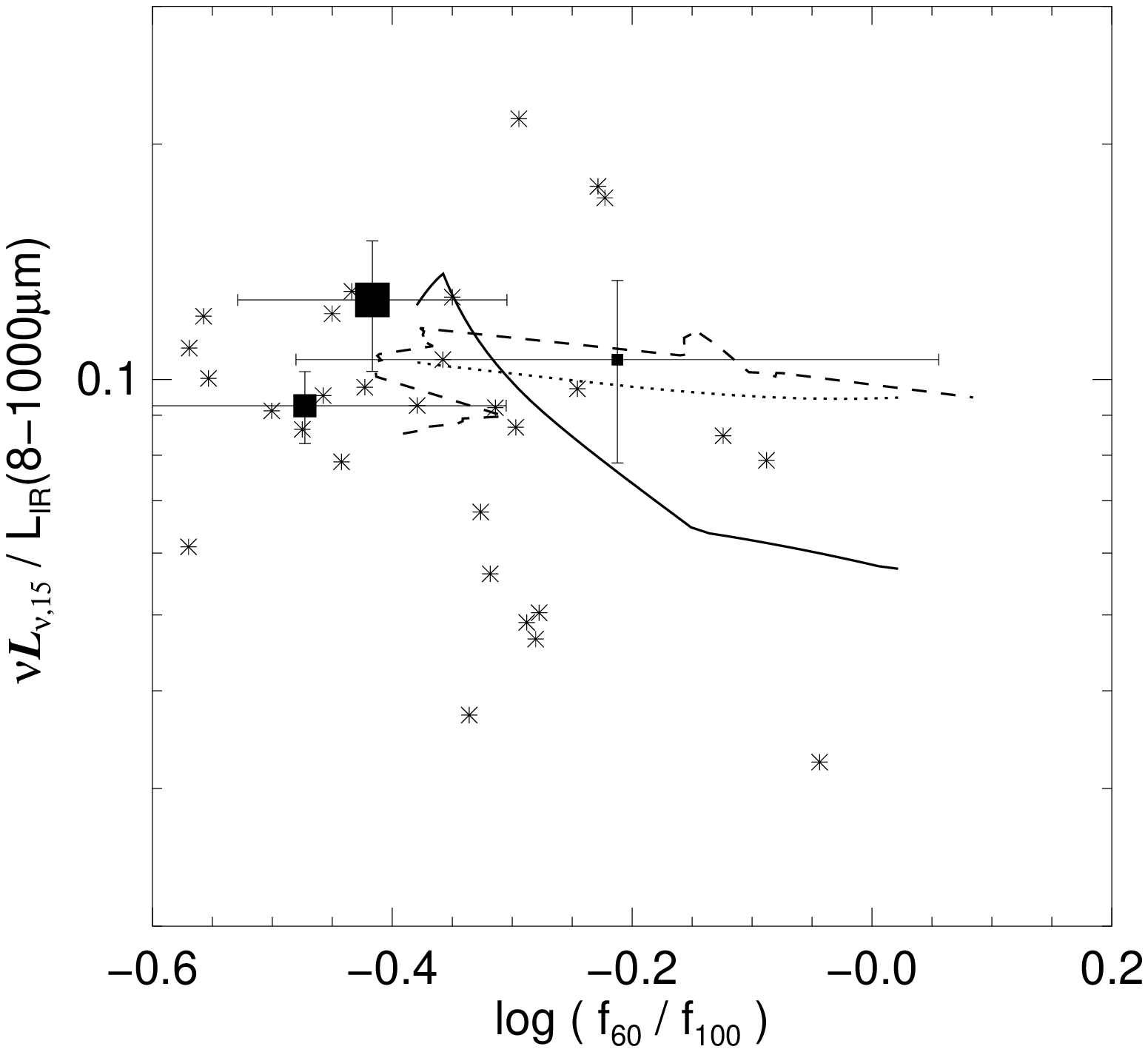}
\caption{The ratio between the rest-frame 15\,$\micron$ to the 12--100\,$\micron$
  luminosity  ({\it left} plot) and to the total IR
  luminosity (8--1000\,$\micron$; {\it right} plot) as a function of the IR flux
  ratio $f_{60}/f_{100}$. {\it Asterisks} are local galaxies  collected 
  from literature with IRAS, ISO 15\,$\micron$ observations and at least one
  observation at wavelengths longer than 100\,$\micron$. {\it Squares} show
  our results for galaxies with $M^\ast\ge 10^{10}$\,$M_\odot$ and 0.6$<z<$0.8.
  The relations derived from SED templates are presented for
  comparison. One can see that the extrapolation of the rest-frame 15\,$\micron$ luminosity based on the three sets of templates give comparable estimates of total IR luminosity within a considerable scatter.
Our data points distribute within the scatter of the
local star-forming galaxies but towards the colder templates. 
Roughly speaking, a factor of 10 is a quite good average bolometric correction to the estimate of the total IR luminosity from the rest-frame 15\,$\micron$ luminosity.
Generally speaking, the use of local templates of  $L_{\rm IR} \leq 10^{10.5}$\,$L_\odot$ for extrapolation of total IR luminosity from intermediate redshift 24\,$\micron$ luminosities gives a more accurate result than the use of  $L_{\rm IR} \sim 10^{11}$\,$L_\odot$ templates. 
}\label{colorlum}
\end{figure*}

\subsection{Comparison with local SEDs}

A key goal of this paper is to compare the observed average IR SED shapes at
$z \sim 0.7$ to local `template' SEDs.  We adopted a sample of 
local star-forming galaxies from \citet{Dale00} with $IRAS$ and ISO 15\,$\micron$ observations and total IR
luminosity (8 - 1000\,$\micron$) spanning from $\sim 10^9$ to
$10^{12}$\,$L_\odot$.  
Figure~\ref{colorcolor} shows these nearby galaxies in the IR flux ratio
$f_{60}/f_{100}$ versus $f_{15}/f_{60}$ plot.
The local galaxies distribute along a sequence with considerable scatter. 
The sequence is correlated with both dust temperature and IR
luminosity.  The dust temperature and IR luminosity increase for increasing
$f_{60}/f_{100}$ \citep[e.g.,][]{Soifer91}.
The average IR SEDs for $z\sim 0.7$ galaxies are determined by the three MIPS
bands, corresponding to the rest-frame $\sim$14, $\sim$41 and
$\sim$94\,$\micron$ bands. The 24\,$\micron$ and 160\,$\micron$
luminosities can be taken as rest-frame 15\,$\micron$ and 100\,$\micron$
luminosities, for which `$K$-corrections' are negligible.
We estimated the rest-frame 60\,$\micron$ luminosity by linear interpolation
between the MIPS measurements in $\log$-$\log$ space (as shown in Figure~\ref{sedsmass}). We used the local sample to test the linear interpolation. First, we derive the 41\,$\micron$ fluxes by linear 
interpolation between $IRAS$ 25 and 60\,$\micron$ measurements in $\log$-$\log$ space. 
Second, we derive 60\,$\micron$ fluxes by the same method between 41 and 
100\,$\micron$ for the 59 local galaxies adopted. 
The estimated 60\,$\micron$ fluxes are 25$\pm$11\% lower than 
the observed $IRAS $ 60\,$\micron$ fluxes. This hints that our 
rest-frame 60\,$\micron$ luminosities of the  $z\sim 0.7$ galaxies might be 
underestimated. 
We compared the $z\sim 0.7$ galaxies to the local galaxies. 
As shown in  Figure~\ref{colorcolor}, the two populations are roughly 
located in the same region of the $f_{60}/f_{100}$ versus $f_{15}/f_{60}$ plane. 
Specifically, the $z\sim 0.7$ star-forming
galaxies of IR luminosity $11< \log (L_{\rm IR}/L_\odot) < 11.4$ (those
in the $L_{24}$-high bin and $L_{24}$-medium bin;  the estimates of the total
IR luminosities will be discussed later) populate the relatively
low-temperature 
end of the template sequence.  In the local Universe, these low-temperature
galaxies tend to be of relatively low luminosity 
$\log (L_{\rm IR}/L_\odot) < 10.5$.
This, with significant uncertainties, suggests 
that the typical dust temperature of $z \sim 0.7$
luminous IR galaxies (LIRGs; i.e., galaxies with $\log (L_{\rm IR}/L_\odot) > 11$)
is lower than that of local galaxies of comparable IR luminosity.
The IR flux ratios are poorly determined for the $L_{24}$-low bin because of
the large uncertainties in the average 70 and 160\,$\micron$ IR luminosities,
which are partially due to the intrinsic scatter among the sample galaxies in 
the $L_{24}$-low bin (including early-type galaxies and late-type galaxies in
the quiescent star formation phase).

We compared our average SEDs with the IR SED model templates
from \citet{Lagache04}, \citet{Chary01} and \citet{Dale02}. 
The three sets of templates were empirically calibrated to represent
local star-forming galaxies spanning a wide range in the IR flux
ratio $f_{60}/f_{100}$. 
Dale \& Helou's IR SED templates are characterized by the flux ratio 
$f_{60}/f_{100}$. We use the equation from \citet{Chapman03} to parameterize the IR
luminosity (8 - 1000\,$\micron$) as a function of $f_{60}/f_{100}$.\footnote{The total IR luminosity defined in \citet{Chapman03} is between 3 and 1100\,$\micron$ , in good
  agreement with the adopted one (\citealt{Dale01}; see also
  \citealt{Takeuchi05b}).}
The templates from \citet{Lagache04} and \citet{Chary01} were characterized by the IR luminosity.
We extended each set of SED templates by linear interpolation in logarithmic
space to a grid of SEDs with the
characteristic IR luminosity ranging from 10$^9$ to 10$^{13}$\,$L_\odot$ with a
resolution of 0.1\,dex.   
The observed IR SED of each subset was compared to the three sets of templates:
\begin{equation}
\chi^2 = \sum_{i=1}^{N_{\rm filters}} \left[{L_{{\rm obs},i}
- f_{scale} \times L_{{\rm temp},i}(z=0.7)\over \sigma_i} \right]^2 \:,
\end{equation}
where $L_{{\rm obs},i}$, $L_{{\rm temp},i}$ and $\sigma_i$ are the observed
and template luminosities and their uncertainty in filter i, respectively, and
$f_{\rm scale}$ is a normalization constant. Here only 24, 70 and 160\,$\micron$ bands
were used to fit templates, i.e., $N_{\rm filters}=3$.  
$L_{\rm temp}$ is calculated by convolving the redshifted SED template to
$z=0.7$ with the 24, 70 or 160\,$\micron$ filter transmission function. 
$f_{\rm scale}$ is chosen to minimize the $\chi^2$ for each
template. The top plot in Figure~\ref{sedsmodel} shows the fitting results. 
The templates with a normalization constant $f_{\rm scale}$ of unity are chosen as the
best-fit luminosity match templates and those with a minimum $\chi^2$ as the
best-fit dust temperature match (or SED shape match) templates.
The best-fit templates are compared to the observed SEDs in
the bottom plot of Figure~\ref{sedsmodel}. 

As shown in Figure~\ref{sedsmodel} the local SED templates track the observed
SEDs reasonably well, in particular for the $L_{24}$-medium
bin, which contains the majority of intense star-forming galaxies.  
As the total IR luminosity is dominated by emission in the
rest-frame wavelength range between 10 and 100\,$\micron$, the estimates of
the total IR luminosity with templates from different models give similar
results. 
In contrast, $\chi^2$ is a measure of the shape agreement between a
template and an observed SED. 
It is worth noting that the estimate of total IR luminosity based on the three
measurements at 24, 70 and 160\,$\micron$ is not sensitive
to the shapes of the IR SED templates. Both the best-fit luminosity match and the 
best-fit dust temperature match the SED templates and suggest a total IR (8 -
1000\,$\micron$) luminosity of $\log L_{\rm  IR}/L_\odot\sim$\,11.4, 11.1,
10.3 for the three average SEDs, respectively. 
Generally speaking, the shape (or dust
temperature) of the average IR SEDs of star-forming galaxies at $z\sim 0.7$
(i.e., the $L_{24}$-high bin and $L_{24}$-medium bin) are better fitted by 
the local SED templates of characteristic IR luminosity $L_{\rm IR} \leq
10^{11}$\,$L_{\odot}$ than those of $L_{\rm IR} > 10^{11}$\,$L_{\odot}$. This
holds for all three sets of templates. 
It confirms that the typical star-forming galaxies at $z\sim 0.7$ are likely
to have relatively colder dust emission than local galaxies with comparable IR
luminosity.

\subsection{The extrapolation from the rest-frame 15\,$\micron$ to the total IR luminosity}\label{24toir}

Local IR SED templates are often used to estimate the total IR
luminosities from single mid-IR band luminosities for distant star-forming
galaxies. With the averaged IR SEDs determined in the three MIPS bands, we
are able to better constrain the estimates of the IR luminosities. 
We compared the estimates with those transformed from single 15\,$\micron$
luminosities using local SED templates. 

By linearly integrating between the average 24, 70 and 160\,$\micron$
luminosities (in logarithm space) for the star-forming galaxies at $z\sim$\,0.7, we
estimated the rest-frame 12--100\,$\micron$ luminosities. 
Uncertainties are calculated from the combination of the uncertainties in the three
bands. The total IR (the rest-frame 8--1000\,$\micron$) luminosity is derived
from the three dust temperature-match templates to each observed IR SED (the 24, 70
and 160\,$\micron$). We combine the scatter between the templates and the
uncertainty in determining the dust temperature-match templates, in which the
uncertainties of the 24, 70 and 160\,$\micron$ luminosities are counted,
as the uncertainty for the total IR luminosity. The results are listed in Table~\ref{table}.
For comparison, we use the NASA/IPAC Extragalactic database (NED) to collect a
sample of 29 local star-forming galaxies ($z<0.1$) with observations
at 12, 25, 60 and 100\,$\micron$ by $IRAS$, at 15\,$\micron$ by ISO and at least
in one band longer than 100\,$\micron$.  The 12--100\,$\micron$ luminosity and
the total IR luminosity are calculated as above. 
When observations do not reach 1000\,$\micron$, a modest ($\la 20$\% of the IR luminosity) 
extrapolation is employed using the local SED templates. 

Figure~\ref{colorlum} shows the extrapolations (bolometric corrections) from the rest-frame 15\,$\micron$
to the 12--100\,$\micron$ luminosity (left plot) and to the total IR luminosity (right
plot) as functions of the IR flux ratio $f_{60}/f_{100}$.  
The local star-forming galaxies exhibit a significant scatter 
(\citealt{Chary01}; see also \citealt{Dale05}); such scatter is typically adopted
as the systematic uncertainty of template-based estimates
of total IR luminosity from rest-frame mid-IR luminosity.
The $z\sim 0.7$ star-forming galaxies distribute within the scatter of the
local star-forming galaxies. 
Again, there is a tendency for the $z \sim 0.7$ galaxies to cluster
towards the colder templates, suggesting that use of spiral
galaxy templates for extrapolation of total IR luminosity from 
intermediate redshift 24\,$\micron$ fluxes gives a more accurate result
than the use of starburst/LIRG templates.

\begin{figure}[] \centering
  \includegraphics[width=0.48\textwidth,clip]{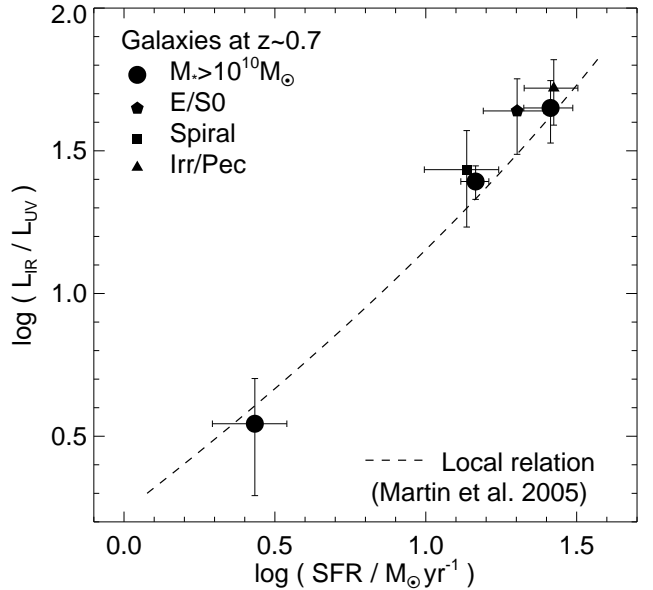}
\caption{The ratio IR/UV flux (i.e., the fractional dust obscuration) as a function of SFR. 
  {\it Solid circles} show the three subsets of galaxies with $M_\ast \ge
  10^{10}$\,$M_\odot$ and $0.6<z<0.8$, sorted by the 24\,$\micron$ luminosity. 
  Other symbols show three morphology subsets of star-forming
  (f$_{24}>83\mu$Jy) galaxies in redshift slice $0.65\leq z\leq 0.75$.  
  The {\it dashed line} shows the local relation
  \citep{Martin05b}.  
}\label{sfrextinction}
\end{figure}

\subsection{Dust extinction}\label{sec:extinc}

A universal relation between SFR and dust extinction is suggested by several
studies at $z < 1$
(\citealt{Hopkins01,Adelberger00,Bell03,Takeuchi05a,Zheng06,Buat06}; although 
see \citealt{Reddy06} for a dissenting view at $z \sim 2$).
With a measured estimate of the total IR luminosity from the 24, 70 and
160\,$\micron$ luminosities, we explore the
relationship between SFR and dust extinction at $z\sim0.7$, comparing it to
that at $z\sim 0$. The dust extinction is described by the IR to UV ratio
$L_{\rm IR}/L_{\rm UV}$.  The UV luminosity, i.e., the integrated luminosity
between rest-frame 1500-2800\,\AA, is estimated from linear
interpolation of the $FUV$, $NUV$, $U$, and $B$ band luminosities for 
$z\sim 0.7$ galaxy subsets shown in Figures~\ref{sedsir} and~\ref{sedsmass}. 
Following \citet{Bell05}, we derived the SFR from the IR and UV
luminosities with the formula
\begin{equation}
{\rm SFR}/(M_\odot\,{\rm yr}^{-1}) = 9.8\times10^{-11}(L_{\rm IR}+2.2L_{\rm UV}),
\end{equation}  
assuming a stellar population with a constant SFR for 100\,Myr and a Kroupa initial mass function.
Figure~\ref{sfrextinction} shows the relationship between dust
obscuration and SFR for $z\sim 0.7$ galaxies, compared to the  local
relation derived from IRAS and GALEX data with a scatter at $\sim$0.5-1\,dex
level \citep{Martin05b,Xu06,Buat06}.
As shown in Figure~\ref{sfrextinction}, the $z\sim 0.7$ galaxies
distribute perfectly along the local relation over one order of magnitude.

\section{Discussion and conclusion} \label{discuss}

We selected a sample of 152 galaxies in the redshift slice $0.65\leq z \leq
0.75$ of known morphology from HST imaging and another sample of 579
mass-limited ($M_\ast\ge 10^{10}$\,$M_\odot$) galaxies in the redshift
slice $0.6<z<0.8$ to study the IR SEDs at that redshift. 
We divided our sample galaxies into different mass-limited morphology and
24\,$\micron$ 
luminosity bins. For each bin, we determined the average luminosities in 14
bands from the FUV to the FIR by summing the individual detections
and adding in the stacked flux from non-detections.   Careful efforts
were taken in stacking the noise and confusion limited 70 and 160\,$\micron$
images. Empirical corrections, determined from the 24\,$\micron$ image, 
were introduced to account for the clustering effects on the stack results.

\begin{figure}[] \centering
  \includegraphics[width=0.48\textwidth,clip]{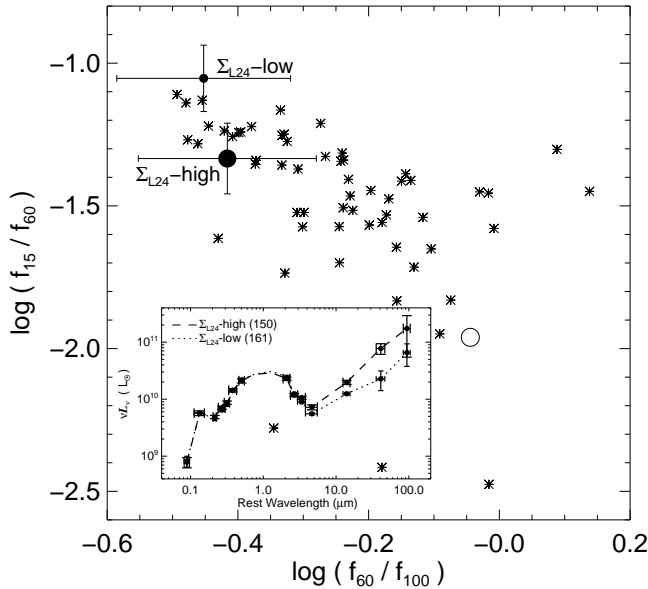}
\caption{Same as Fig.~\ref{colorcolor}.
The {\it filled  circles} show the two subsets of galaxies in the redshift
slice $0.6<z<0.8$ split by 24\,$\micron$ luminosity surface density. The inner
panel shows the average SEDs of the two subsets. The $\Sigma_{\rm L24}$-high
bin and the $\Sigma_{\rm L24}$-low bin data points are located separated along
the dust-temperature sequence of local star-forming galaxies, indicating that
the $\Sigma_{\rm L24}$-high bin is characterized by a hotter dust emission
than the $\Sigma_{\rm L24}$-low bin. This is clearly seen from the difference
between their IR SEDs.
}\label{sfrd}
\end{figure}

The average luminosities in three MIPS bands determine the IR SED from the
rest-frame 10 to 100\,$\micron$. 
Our principal result is that the average IR SED shape of 
$z \sim 0.7$ intensely star-forming
galaxies (with IR luminosities $\sim 10^{11} L_{\sun}$) 
is similar to reasonably `cool' local templates (i.e., templates
of `normal' spiral galaxies).  The dust SED seems 
to depend on morphological type for star-forming
galaxies at $z \sim 0.7$. 
This has the immediate and important
implication that the use of local templates to extrapolate
total IR luminosity from observed-frame 24\,$\micron$
data is a well-posed problem, at least, on average.  
Interestingly, galaxies with  
`cool' dust temperatures in the local Universe all tend to have
IR luminosities $\la 10^{10.5} L_{\sun}$, i.e., distant intensely
star-forming galaxies tend to be characterized by colder dust
emission than their local counterparts of comparable IR luminosity.

Previous studies have found evidence for a somewhat cooler
dust SED at $0.2<z<2.5$ than for local galaxies of a comparable luminosity 
\citep[e.g.,][]{Pope06,Sajina06} at $0.2<z<2.5$.  
These studies were selected in rest-frame $>100$\,$\micron$ 
emission, and the authors suspected that their overall tendency
towards `colder' IR SEDs was in part due to that long-wavelength 
selection.  Our sample is selected on rest-frame $\sim 15$\,$\micron$ 
emission, i.e., by warm dust; yet, we find a `colder' average
SED at a given luminosity than is found locally.  This tends
to support the interpretation that the offset which we and others 
have found towards colder temperatures at a given luminosity are
at least in part a real difference.  We suggest that this tendency toward
colder dust temperature reflects a difference in dust and star
formation geometry: whereas local LIRGs tend to be interacting
systems with relatively compact very intense star formation and of comparable
masses \citep[e.g.,][]{Wang06},
$0.5\la z \la 1$ LIRGs tend to be disk-dominated, relatively
undisturbed galaxies \citep{Zheng04,Bell05,Melbourne05}.
We suggest that these disk galaxies host widespread intense star
formation, like star formation in local spirals but scaled up, 
leading to relatively cold dust temperatures. This is consistent with what
\citet{Chanial06} found --- the scatter in the $L_{\rm
  IR} - T_{\rm dust}$ relation for star-forming galaxies is largely induced by
the size dispersion of the star-forming regions; and the IR luminosity surface
density $\Sigma_{\rm IR} - T_{\rm dust}$ relation is more fundamental than the
the $L_{\rm IR} - T_{\rm dust}$ relation.  We examined the IR flux ratio
$f_{60}/f_{100}$ (indicator of dust temperature) as a function of the IR
luminosity surface density for $z\sim 0.7$ star-forming galaxies. By taking 
galaxy size (i.e. half-light radius) derived from HST imaging \citep{Haussler07} as a proxy of the size of star-forming regions
in a galaxy and 24\,$\micron$ luminosity as a proxy of total IR
luminosity, we split a sample of 311 24\,$\micron$-detected galaxies in the
redshift slice $0.6<z<0.8$ into two bins in the 24-$\micron$ luminosity
surface density $\Sigma_{\rm L24}$. The average SEDs for the two subsets of
galaxies were constructed in the same way as described in \S\ref{consed} and
shown in the inner panel of Figure~\ref{sfrd}. The total IR luminosity is
estimated as $\log (L_{\rm IR}/L_\odot) = 11.3$ and 11.0 for the 
$\Sigma_{\rm L24}$-high bin and the $\Sigma_{\rm L24}$-low bin respectively. 
As shown in Figure~\ref{sfrd}, the two subpopulations are
distributed along the dust-temperature sequence of local
star-forming galaxies. Galaxies in the $\Sigma_{\rm L24}$-high bin
on average show hotter dust emission than those in the $\Sigma_{\rm L24}$-low bin, 
suggesting that the IR luminosity surface density plays an essential role in
shaping IR SED \citep{Chanial06}.

Finally, with UV luminosities derived 
from GALEX data and IR luminosities derived from
three band MIPS data, we determined a quite precise relationship
between the dust obscuration and SFR for $z\sim$0.7. 
Our results show an excellent agreement between the SFR-dust obscuration
relation at $z\sim$0.7 with that at the present day, indicating that no significant
evolution occurs since that redshift. Our measurements give the
mean values of SFR and dust obscuration. The actual scatter in dust
obscuration  can spread by 1 - 2\,dex for individual galaxies of given
SFR\citep{Martin05b,Xu06,Buat06}. 
\citet{Reddy06} claimed that the dust obscuration for
star-forming galaxies is systematically smaller at $z \sim 2$ than the present
day. Their sample is dominated by rest-frame UV-selected galaxies. 
This may lead to a potential selection bias against objects with
high IR/UV ratio (see \citealt{Buat06} for the comparison
between UV and FIR selected samples of local galaxies and a similar
discussion).  
Therefore studies based on unbiased samples of high-redshift galaxies will
help to answer the question whether the SFR-dust obscuration relation still
holds at $z>1$. 
This will add important constraints to our understanding of galaxy evolution
involving star formation and metallicity enrichment \citep{Zheng06}.

\begin{acknowledgements}
  
  We are grateful to Daniel A. Dale and Kirsten K. Knudsen for helpful
  discussions. E.\ F.\ B.\ was supported by
  the Emmy Noether Programme of the Deutsche Forschungsgemeinschaft.
  Support for E.\ L.\ F.'s work was provided by NASA through the Spitzer Space 
  Telescope Fellowship Program. This work was in part supported by
contract 1255094 from JPL/Caltech to the University of Arizona.
  This research has made use of the NASA/IPAC Extragalactic Database (NED) which is operated by the Jet Propulsion Laboratory, California Institute of Technology, under contract with the National Aeronautics and Space Administration.
\end{acknowledgements}

%%%%%%%%%%%%%%%%%%%%%%%%%

%%% Local Variables: 
%%% mode: latex
%%% TeX-master: "ms_z0.7seds1"
%%% End: 

\end{document}